\documentclass[
    aps,
    prx,
    reprint,
    superscriptaddress,
    longbibliography,
    amsmath,
    amssymb,
    floatfix
]{revtex4-2}

\usepackage{graphicx}
\usepackage{indentfirst}
\usepackage{siunitx}
\usepackage{physics}
\usepackage{braket}
\usepackage{float}
\usepackage{mathtools}
\usepackage{epstopdf}
\usepackage{esint}
\usepackage{comment}
\usepackage{color}
\usepackage[T1]{fontenc}
\usepackage{footmisc}
\usepackage{scrextend}
\usepackage{multirow}
\usepackage[hyperfootnotes=false]{hyperref}
\usepackage[acronym]{glossaries}
\usepackage[english]{babel}
\usepackage{url}
\usepackage{bm}
\usepackage{tikz}
\usepackage{etoolbox}
\usepackage{enumerate}
\usepackage{soul}
\usepackage{amssymb}
\usepackage[caption=false]{subfig}
\definecolor{darkblue}{rgb}{0,0,0.5}
\hypersetup{
    colorlinks=true,
    linkcolor=black,
    filecolor=blue,
    citecolor=darkblue,  
    urlcolor=black,
}

\usepackage[normalem]{ulem}
\usepackage{algorithm} 
\usepackage{algpseudocode} 

\newcommand{\calA}{\mathcal{A}}

\newcommand{\calD}{\mathcal{D}}
\newcommand{\calE}{\mathcal{E}}

\newcommand{\calG}{\mathcal{G}}
\newcommand{\calH}{\mathcal{H}}

\newcommand{\calO}{\mathcal{O}}

\newcommand{\calS}{\mathcal{S}}
\newcommand{\calT}{\mathcal{T}}

\newcommand{\state}[1]{\ketbra{#1}{#1}}
\newcommand{\roundket}[1]{\left|{#1}\right)}
\newcommand{\roundbra}[1]{\left({#1}\right|}

\newcommand{\bI}{\boldsymbol I}

\newcommand{\bmz}{{\bm z}}

\def\be{\begin{equation}}
\def\ee{\end{equation}}
\def\ba{\begin{eqnarray}}
\def\ea{\end{eqnarray}}

\apptocmd{\sloppy}{\hbadness 9999\relax}{}{}

\DeclareMathOperator{\E}{\mathbb{E}}

\DeclarePairedDelimiterX\parvertent[2]{\lparen}{\rparen}%
{#1\,\delimsize\vert\,\mathopen{}#2}

\newtheorem{theorem}{Theorem}
\hypersetup{colorlinks=true,citecolor=blue,linkcolor=blue,urlcolor=blue}

\begin{document}

\title{Measurement-induced overconcentration in quantum generative models}

% Replace the placeholders below with your actual author list.
\author{Runzhe Mo}
\email{runzhemo@usc.edu}
\affiliation{Ming Hsieh Department of Electrical and Computer Engineering, University of Southern California, Los Angeles, California 90089, USA}
\author{Bingzhi Zhang}
\affiliation{Ming Hsieh Department of Electrical and Computer Engineering, University of Southern California, Los Angeles, California 90089, USA}
\author{Quntao Zhuang}
\email{qzhuang@usc.edu}
\affiliation{Ming Hsieh Department of Electrical and Computer Engineering, University of Southern California, Los Angeles, California 90089, USA}
\affiliation{Department of Physics and Astronomy, University of Southern California, Los Angeles, California 90089, USA}

\begin{abstract}

Quantum measurement is a key resource for quantum generative learning, providing intrinsic stochasticity for generating diverse quantum samples. However, in measurement-assisted state-ensemble resampling, repeated measurements can also induce overconcentration: under a fixed measurement trajectory, distinct input states progressively converge toward similar output states, suppressing input-dependent diversity. To diagnose this effect, we introduce three complementary metrics: accuracy, generative power, and input sensitivity. For Haar-random monitored circuits, we prove that one-step models retain input sensitivity up to dimension-suppressed corrections, whereas sequential monitored circuits exhibit a depth-dependent loss of input sensitivity. Motivated by this diagnosis, we propose a truncated quantum denoising diffusion probabilistic model (QuDDPM), which restricts the temporal depth of both the forward diffusion and reverse denoising processes. Numerical benchmarks show that truncated QuDDPM preserves stronger input sensitivity while maintaining accuracy and generative power comparable to the original model. These results identify measurement-induced overconcentration as a dynamical limitation of deep monitored quantum generative models and establish temporal depth as a design parameter for balancing measurement-induced randomness with input-dependent diversity.

%Quantum measurement plays a central role in quantum generative learning, yet its role in inference tasks such as state-ensemble resampling is not fully characterized by target-ensemble matching alone. To evaluate measurement-assisted inference, we establish three complementary metrics: accuracy, generative power, and input sensitivity. These metrics quantify target matching, generation from limited input resources, and preservation of input-ensemble structure in conditional outputs. Applying this framework, we find that one-step measurement-assisted models retain input sensitivity but have limited accuracy and generative power, whereas sequential models improve generation performance while progressively suppressing input sensitivity. This loss arises from repeated measurement-induced conditioning: for a fixed measurement trajectory, distinct input states increasingly concentrate toward similar output states. Guided by this diagnosis, we propose a truncated quantum denoising diffusion probabilistic model (QuDDPM) that restricts temporal depth. Numerical benchmarks on circular states in logical space show that truncated QuDDPM preserves stronger input sensitivity while maintaining accuracy and generative power comparable to the original model. These results identify a trade-off in measurement-assisted quantum generation and establish performance metrics for quantum generative models beyond distribution matching.

\end{abstract}

% Optional: uncomment if you want displayed keywords with the showkeys class option.
% \keywords{quantum generative learning, diffusion models, monitored circuits, mid-circuit measurement}

\maketitle

% ==========================================================
% PRX submission note (not part of the manuscript body)
% ==========================================================
% Physical Review X requires a nontechnical popular summary at submission.
% That summary is usually supplied in the APS submission system rather than
% embedded in the main manuscript source. Draft one separately in plain language.
%
% Suggested one-sentence draft:
% We show that mid-circuit measurements can make quantum generative models
% overly concentrated, and we introduce a truncated diffusion-style method
% that preserves more diversity while maintaining good generation quality.

\section{Introduction}
\label{sec: intro}

Generative learning aims to model complex data distributions and produce new samples from them. In classical machine learning, this goal underlies variational autoencoders \cite{Kingma2014Auto-Encoding}, autoregressive models \cite{oord2016pixel}, generative adversarial networks \cite{goodfellow2014generative}, normalizing flows \cite{Rezende2015Variational}, score-based diffusion models \cite{ho2020denoising,song2021scorebased}, and flow-matching approaches \cite{Lipman2023Flow}. 
The quantum analogue seeks to learn quantum processes that generate samples reproducing the structure of target data, either encoded in measurement statistics or represented directly as quantum states and state ensembles, as explored in quantum circuit Born machines \cite{liu2018differentiable,coyle2020born}, quantum generative adversarial networks \cite{lloyd2018quantum,dallaire2018quantum}, and quantum diffusion models \cite{zhang2024generative, parigi2025quantum}. To apply these generative models to the quantum state resampling task, subsystem measurement becomes a natural resource---resulting in a framework of measurement-assisted generative learning, where single-step partial measurement or multiple-step mid-circuit measurement can be applied. In this measurement-assisted generative model, a central question is how different measurement schemes affect the generative performance of the model.

In this work, we reveal a measurement-induced overconcentration phenomenon in measurement-assisted quantum generative models. While multiple steps of mid-circuit measurements enhance the generative power and accuracy, it inevitably leads to the waste of input-state variation---distinct input states can converge toward similar output states under a fixed measurement trajectory. To quantitatively uncover the overconcentration phenomenon, we introduce quantifiers for the three complementary metrics: accuracy, generative power, and input sensitivity, noticing that existing criteria~\cite{du2022efficient,sim2019expressibility,mcclean2018barren,cerezo2021cost,zhang2025energy,lu2020quantum,liao2021robust} are insufficient for our purpose. In particular, input sensitivity captures the extent to which the input ensemble structure is preserved in the conditional generated ensemble.  These metrics provide a model-agnostic language for analyzing measurement-assisted quantum generative models beyond distribution matching alone.

For Haar-random monitored circuits, we show that, as the number of temporal steps increases, input sensitivity decays linearly with the number of steps. While this scaling is derived for Haar-random monitored circuits, we further observe the same qualitative concentration behavior in optimized quantum denoising diffusion probabilistic model (QuDDPM)~\cite{zhang2024generative, kwun2025mixed}. Overconcentration is detrimental to input-efficient ensemble resampling, since it limits the model’s
ability to convert input variation into output diversity. To mitigate this overconcentration, we propose a truncated QuDDPM protocol that restricts the overall temporal depth of both the forward diffusion and reverse denoising processes. Our numerical results show that truncated QuDDPM provides a better trade-off between the performance metrics. This example illustrates how diagnosing overconcentration can guide the choice of operating regimes in quantum generative learning.

\begin{figure*}[t]
    \includegraphics[width=\textwidth]{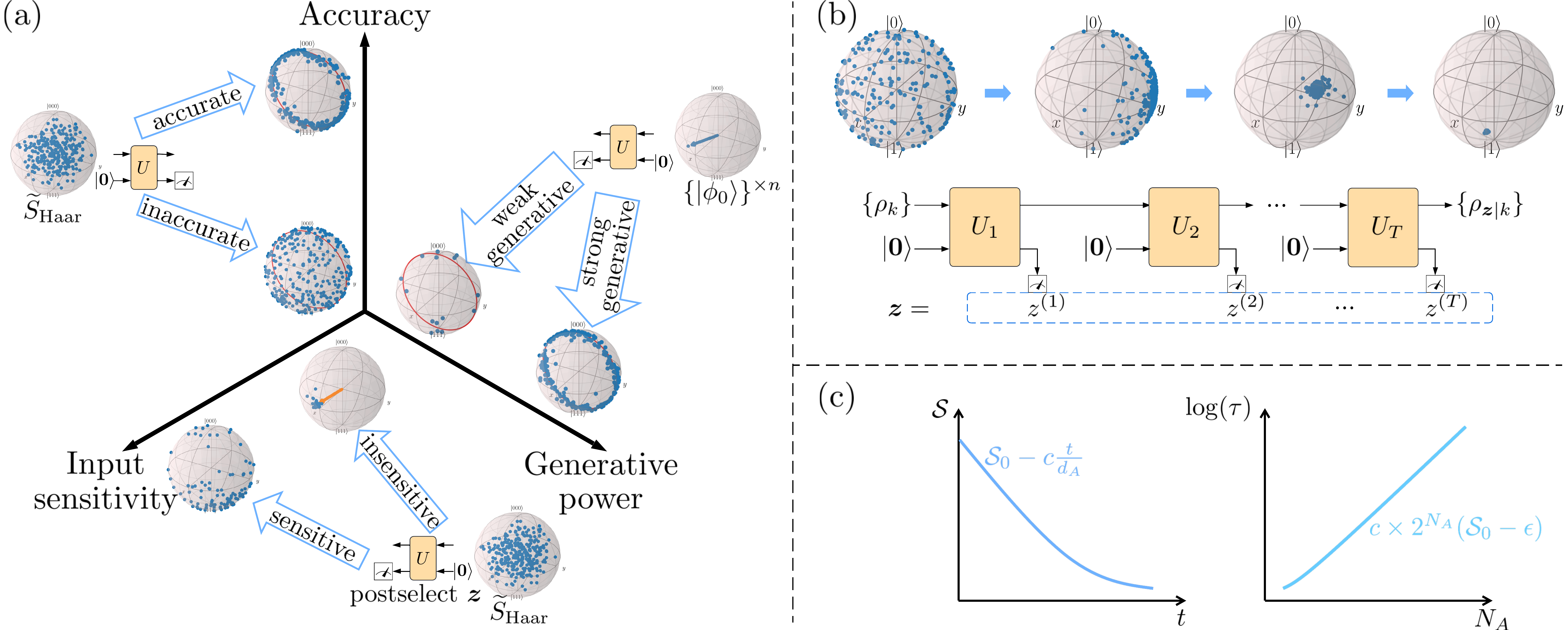}
    \centering
  \caption{
Performance characterization and measurement-induced concentration in measurement-assisted quantum generative learning.
(a) Visualization of three metrics characterizing the performance: accuracy measures the closeness of the unconditional generated ensemble $\mathcal{E}{\rm gen}$ to the target ensemble, generative power measures the generation performance using limited input resources, i.e., a single input state, and input sensitivity measures how much of the input-state ensemble structure is retained in the conditional generated ensemble $\mathcal{E}{{\rm gen}|\bm z}$.
(b) Schematic of measurement-induced concentration in a multistep sequential model. Conditioned on the same measurement trajectory $\bm z$, distinct input states progressively converge toward similar conditional output states as the temporal depth increases.
(c) Scaling behavior of input sensitivity $\calS$ with temporal depth and system size. Input sensitivity initially decays linearly with time before saturating at late times, while the concentration time $\tau$ required to reach a prescribed sensitivity level $1-\epsilon$ scales exponentially with the system size.
}
    \label{fig:overall_scheme}
\end{figure*}

\section{Overview}

Our major contributions are fourfold. First, we establish a generative learning framework that incorporates existing models in a unified fashion.  We consider a measurement-assisted model, in which an input state and a trivially initialized ancilla evolve under a global parameterized unitary, followed by measurement of the ancilla. This setup captures both the trainable unitary transformation and the intrinsic randomness supplied by measurements. It encompasses one-step generative models such as quantum direct transport (QuDT), a variant of Quantum Circuit Born Machines (QCBM)~\cite{liu2018differentiable}, and Quantum Generative Adversarial Network (QuGAN)~\cite{lloyd2018quantum,dallaire2018quantum}, as well as multistep sequential models. Among the latter, the quantum denoising diffusion probabilistic model (QuDDPM)~\cite{zhang2024generative, kwun2025mixed} is particularly appealing because its divide-and-conquer structure enables efficient learning of state ensembles.

Second, we establish three complementary performance metrics to evaluate the performance of general generative models (see Fig.~\ref{fig:overall_scheme}a). General quantum machine learning diagnostics, including expressivity~\cite{du2022efficient,sim2019expressibility}, trainability~\cite{mcclean2018barren,cerezo2021cost,zhang2025energy}, and robustness~\cite{lu2020quantum,liao2021robust}, characterize important properties of variational models but do not directly capture ensemble-resampling performance.
To probe this performance, we propose three complementary metrics: accuracy, generative power, and input sensitivity. 
%Accuracy quantifies how close the generated ensemble is to the target ensemble with input states sampled from a prior distribution. Generative power characterizes how well the target ensemble can be reproduced from highly limited input resources, i.e., a single state. Input sensitivity captures the extent to which the input ensemble structure is preserved in the conditional generated ensemble. 
As shown in Fig.~\ref{fig:overall_scheme}a, while accuracy captures whether the generated ensemble matches the target (e.g., a red circle on the Bloch sphere in Fig.~\ref{fig:overall_scheme}a), input sensitivity describes whether the input ensemble randomness is discarded (insensitive) or preserved (sensitive), and generative power represents how much randomness the quantum generative model is able to generate from a single input state. 

Third, in terms of input sensitivity, we uncover an overconcentration phenomenon in measurement-assisted generative models---the generated states tend to concentrate as mid-circuit measurement steps increase (see Fig.~\ref{fig:overall_scheme}b). Using input sensitivity as the diagnostic, we derive its analytical scaling with respect to both system dimension and temporal depth for Haar-random monitored circuits (see Fig.~\ref{fig:overall_scheme}c). 
For one-step models, the loss of input sensitivity scales inversely with the Hilbert-space dimension of the system, so the conditional output ensemble largely preserves the structure of the input ensemble. In contrast, sequential monitored circuits accumulate this loss approximately linearly with temporal depth before saturation, progressively suppressing input dependence through repeated mid-circuit measurements.

Fourth, we propose truncated QuDDPM. Instead of learning a full reverse process from Haar-random states, truncated QuDDPM starts from partially scrambled states and learns a shortened denoising process, reducing the depth over which concentration can accumulate. This truncation can also avoid unnecessary training difficulty induced by highly entangled input states~\cite{marrero2021entanglement,cao2025mitigating}. We benchmark truncated QuDDPM against the original QuDDPM and QuDT for generative learning of a circular state ensemble in logical space. Compared to the original QuDDPM, truncated QuDDPM achieves comparable accuracy and generative power while exhibiting stronger input sensitivity with a reduced temporal depth.

The remainder of the paper is organized as follows. In Sec.~\ref{sec:general}, we introduce the measurement-assisted quantum generative learning framework for ensemble resampling, define conditional and unconditional ensemble transport, and discuss representative quantum generative models within this framework. In Sec.~\ref{sec:metrics}, we introduce three performance metrics for quantum generative models in ensemble resampling. In Sec.~\ref{sec:concentration}, we analyze the concentration phenomenon induced by repeated measurements and derive the scaling of input sensitivity for monitored circuits. In Sec.~\ref{sec:metric_tradeoff_diagnosis}, we use the proposed metrics to diagnose the performance trade-off in the original QuDDPM, together with the one-step QuDT for comparison. In Sec.~\ref{sec:truncate_alt}, we introduce and benchmark truncated QuDDPM, showing that it provides a more favorable operating regime than the original QuDDPM. We conclude in Sec.~\ref{sec:discussion} with implications and open directions.

\section{Quantum generative learning model framework for ensemble resampling}
\label{sec:general}

% \begin{figure}[t]
%     \centering
%     \subfloat[Monitored Circuit\label{monitored_circuit}]{
%         \includegraphics[width=0.95\columnwidth]{fig/monitored_circuit.png}
%     }

%     \vspace{0.1em}

%     \subfloat[Performance Metrics \RM{**Need further polishing}\label{performance_metrics}]{
%         \includegraphics[width=0.95\columnwidth]{fig/metrics.png}
%     }

%     \caption{Two vertically stacked subfigures. \QZ{only general circuit, move monitored circuit to later section}}
%     \label{fig:main}
% \end{figure}

\subsection{Measurement-assisted quantum generative learning paradigm}
\label{subsec:inference_circuit}

\begin{figure}[t]
    \includegraphics[width=0.7\columnwidth]{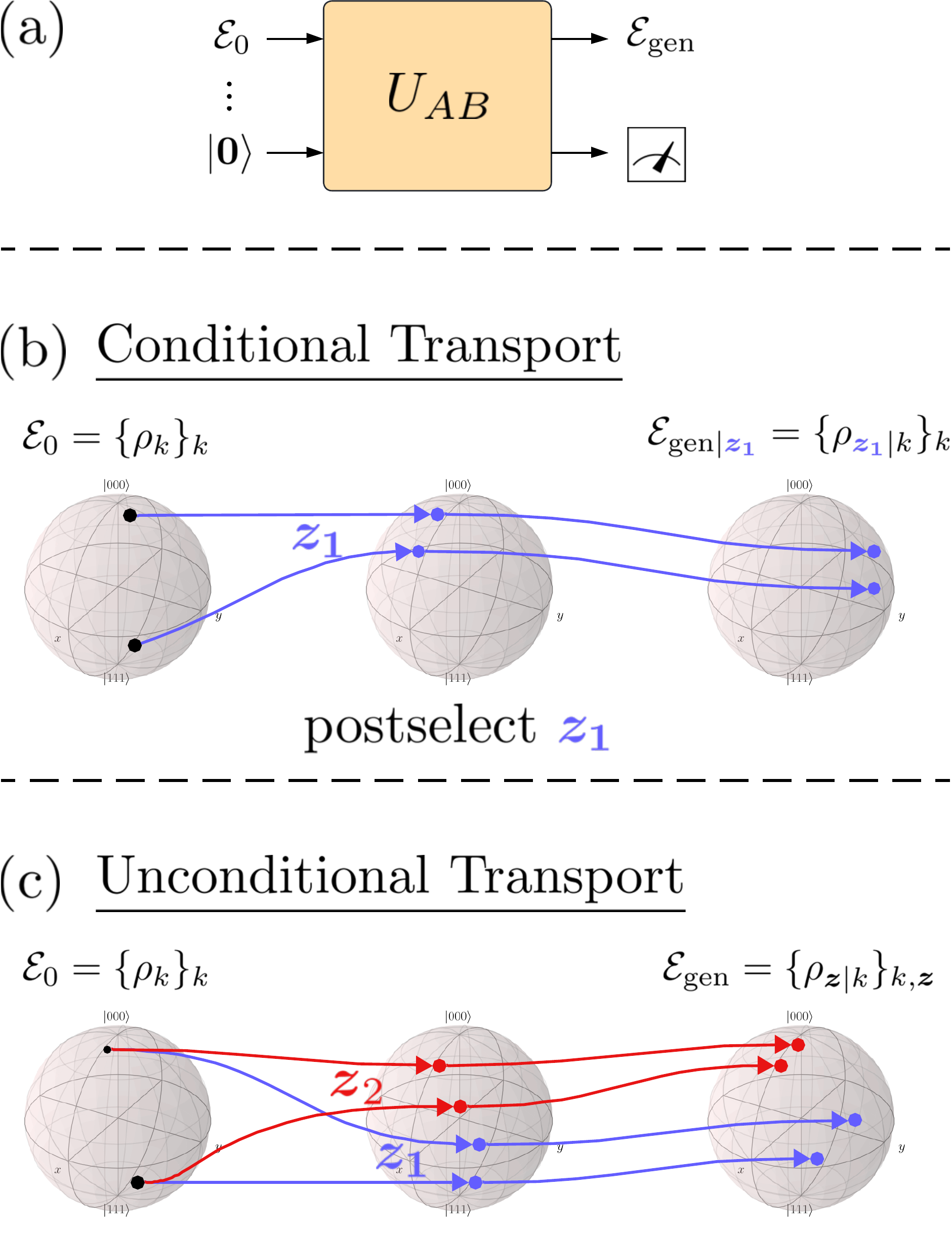}
    % \centering
    \caption{
Measurement-assisted quantum generative model and two transport types. 
(a) An input state from $\mathcal{E}_0$ is coupled to an initialized ancilla by a joint unitary $U_{AB}$, followed by ancilla measurement to produce $\mathcal{E}_{\rm gen}$.
(b) Conditional transport maps different inputs to outputs with a fixed measurement outcome $\bmz_1$. 
(c) Unconditional transport collects all generated states on the output side over all outcomes $\bmz_1, \bmz_2$.
}
\label{fig:framework_transport}
\end{figure}

In this section, we introduce the measurement-assisted quantum generative learning paradigm for a particular task, {\em quantum state ensemble resampling}. In this task, we aim to learn the inherent distribution of a target quantum state ensemble from a finite set of samples and generate new data samples from this ensemble using given input states. As depicted in Fig.~\ref{fig:framework_transport}a, the measurement-assisted model starts from an initial state ensemble $\calE_0$ of $N_A$-qubit states, where we sample a state $\rho \in \calE_0$ to initialize the system $A$. We apply a joint unitary $U_{AB}$ to the system $A$ and an $N_B$-qubit ancilla $B$, which is initialized in the reference state $\ket{\bm 0}_B \coloneqq \ket{0}_B^{\otimes N_B}$, followed by projective measurements in a fixed basis on the ancilla qubits. 
% Conditioned on the measurement outcome $\bmz$, we obtain the output state on the system side as
% \be
%     \rho_\bmz = K_\bmz \rho K_\bmz^\dagger / p_\bmz,
%     \label{eq:cond_state_update}
% \ee
% where $K_\bmz = {}_B\braket{\bmz|U_{AB}|\bm 0}_B$ is the Kraus operator and $p_\bmz = \tr(K_\bmz \rho K_\bmz^\dagger)$ is the outcome probability. 
% Then for every state $\rho_k$ in the input state ensemble $\calE_0=\{p_k,\rho_k\}_k$, the output state conditioned on the measurement outcome $\bmz$ is thus
% \begin{equation}
%     \rho_{\bmz|k}= K_\bmz\rho_k K_\bmz^\dagger / p_{\bmz|k}, 
% \end{equation}
% with outcome probability $p_{\bmz|k}= \tr\left(K_\bmz\rho_k K_\bmz^\dagger \right)$. \BZ{Eq 1 and 2 are kind of similar, shall we keep only one inline equation}
For each measurement outcome $\bm z$, the monitored circuit induces the Kraus operator $K_{\bm z} = {}_B\!\braket{\bm z|U_{AB}|\bm 0}_B$. Acting on each state $\rho_k$ in the input ensemble $\calE_0=\{p_k,\rho_k\}_k$, this gives the conditional output state
\be
    \rho_{\bm z|k} = \frac{K_{\bm z}\rho_k K_{\bm z}^\dagger}{p_{\bm z|k}}, \qquad p_{\bm z|k} = \tr\!\left(K_{\bm z}\rho_k K_{\bm z}^\dagger\right),
    \label{eq}
\ee
where $p_{\bm z|k}$ is the probability of obtaining outcome $\bm z$ given the input state $\rho_k$.
Conditioned on the same measurement outcome $\bmz$, we define the measurement-conditioned generated state ensemble as
\begin{equation}
    \calE_{{\rm gen}|\bmz}\coloneqq \left\{\frac{p_{\bmz|k}p_k}{p_\bmz^{\calE_0}}, \rho_{\bmz|k}\right\}_k,
    \label{eq:cond_ensemble_def}
\end{equation}
where $p_\bmz^{\calE_0}= \sum_k p_k p_{\bmz|k}$ is expected probability of measurement outcome $\bmz$ over the input ensemble. We can thus formulate the transport from the input state ensemble $\calE_0$ to the conditional state ensemble $\calE_{{\rm gen}|\bmz}$ as the {\em conditional transport}. In the conditional transport, only the initial input state varies, while the unitary interaction and measurement outcome are fixed.
% We introduce this conditional transport to characterize the sensitivity of the generated state ensemble to different input states while holding all variables fixed, including unitary and measurement outcomes. 
We visualize this transport in Fig.~\ref{fig:framework_transport}b using the Bloch-sphere representation, where the two blue arrows represent the transport from input to output (left to right) under the same unitary $U$ and the same conditioned measurement outcome $\bmz_1$. Since quantum measurements produce outcomes according to the Born rule, the conditional transport corresponds to the branch conditioned on the observed measurement outcome, as highlighted in Fig.~\ref{fig:framework_transport}b.

In practice, one does not need to condition on a particular measurement outcome on the output side; instead, we collect all the generated states. This leads to the measurement-unconditioned generated state ensemble
\be
    \calE_{\rm gen} = \{p_{\bmz|k}p_k, \rho_{\bmz|k}\}_{k,z},
    \label{eq:uncond_ensemble_def}
\ee
which is obtained by aggregating the conditional ensembles $\calE{{\rm gen}|\bmz}$ over all measurement outcomes, weighted by their occurrence probabilities. We can also formulate the mapping from input state ensemble $\calE_0$ to output state ensemble $\calE_{\rm gen}$ as the unconditional transport, which is visualized in Fig.~\ref{fig:framework_transport}c. Here, blue and red arrows represent different conditional transports corresponding to measurement outcomes $\bmz_1$ and $\bmz_2$, respectively, while arrows with the same color represent conditional states generated from different inputs but the same outcome. 

Finally, we would like to emphasize the importance of the ancilla measurement for state-ensemble resampling. Consider an arbitrary pair of pure states $\ket{\psi_1}, \ket{\psi_2}$ sampled from the input state ensemble $\calE_0$. Without introducing an ancilla system and performing measurements on it, the fidelity of evolved states under a fixed unitary $U$ remains the same $|\braket{\psi_1|U^\dagger U|\psi_2}|^2 = |\braket{\psi_1|\psi_2}|^2$. In other words, the distance between any two inputs is unchanged, and so is the structure of the state ensemble. Therefore, we do not regard a pure unitary without measurement as a model for state ensemble resampling. Moreover, quantum measurements provide randomness in addition to input variation, thus enabling the generation of many more data samples from fewer input samples. Indeed, measurement-assisted circuits have also been adopted recently for random quantum state generation, in both deep thermalization~\cite{ho2022exact} and holographic deep thermalization~\cite{zhang2025holographic}.

\subsection{Current quantum generative models}

\begin{figure}[t]
    \includegraphics[width=\columnwidth]{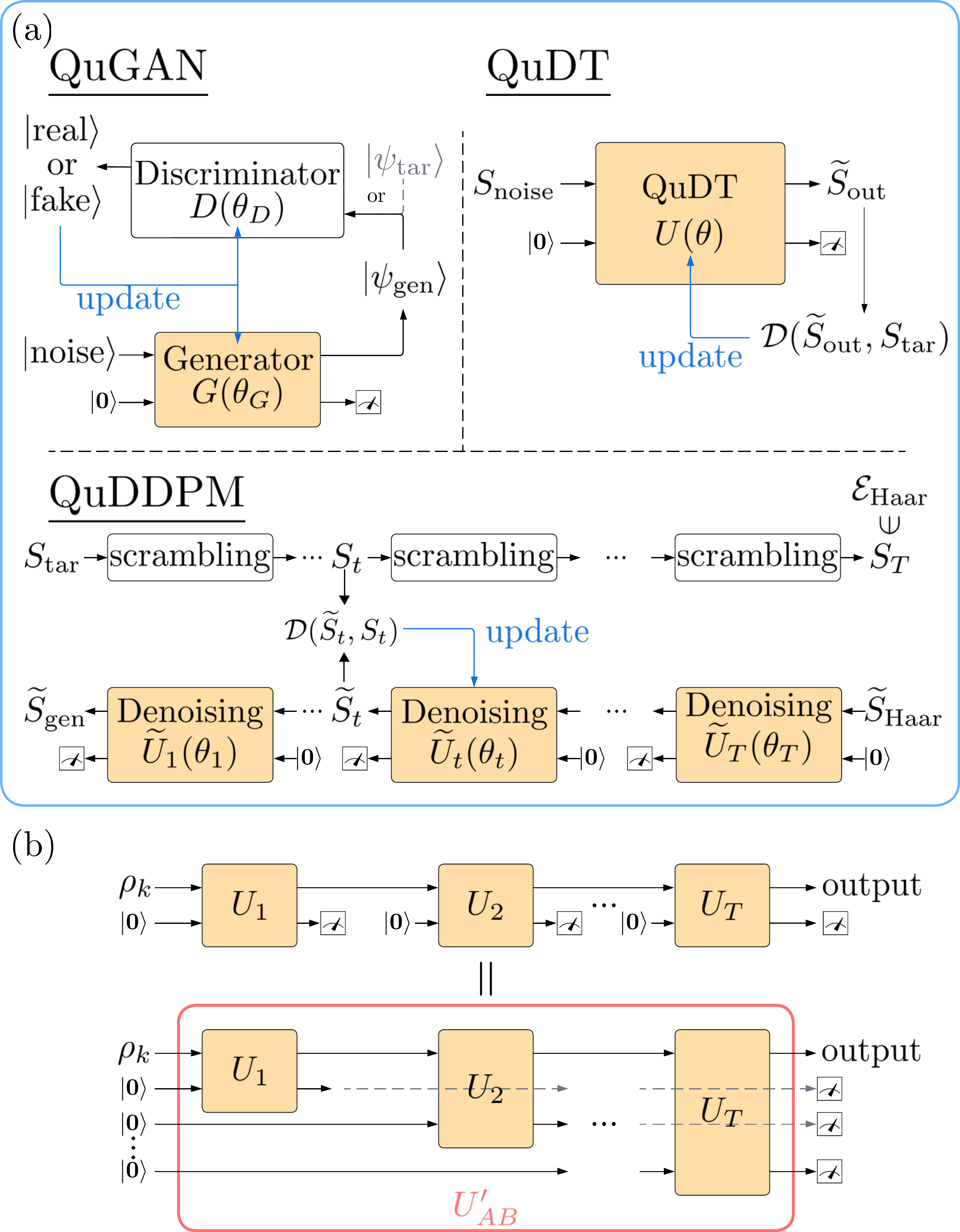}
    % \centering
    \caption{
General and sequential measurement-assisted quantum generative models.
(a) Schematics of Quantum generative adversarial network (QuGAN), Quantum direct transport (QuDT) and Quantum denoising diffusion probabilistic model (QuDDPM).
(b) A sequential monitored circuit can be viewed as a structured realization of the general measurement-assisted model, where a large effective measurement trajectory space is generated by repeatedly reusing a small ancilla register. 
}
\label{fig:general_sequential}
\end{figure}

Within this measurement-assisted transport picture, we now identify several representative quantum generative models analyzed in this work, as summarized in Fig.~\ref{fig:general_sequential}.

{\em QuDT and QuGAN.---} Within this framework, QuDT can be understood as an extension of the known quantum generative model, quantum circuit born machine (QCBM). The original QCBM~\cite{liu2018differentiable} uses a parametrized circuit acting on a fixed input $\ket{0}^{\otimes n}$, and is trained to reproduce a classical bitstring target distribution. QuDT generalizes QCBM in two directions: it replaces the fixed input with an input ensemble $\mathcal{E}_0$, and also introduces ancilla measurements for ensemble learning~\cite{zhang2024generative}. 
% This allows the model to learn a transport map from $\mathcal{E}_0$ to an output ensemble $\mathcal{E}_{\mathrm{gen}}$ approximating the target ensemble $\mathcal{E}_{\mathrm{tar}}$, thereby enabling ensemble resampling. 
Similarly, QuGAN~\cite{lloyd2018quantum, dallaire2018quantum} uses a parameterized quantum circuit as the generator, which is trained adversarially against a discriminator that learns to distinguish generated samples from target samples. The generator of QuGAN can also be modified in the same way as QuDT to accommodate the task of ensemble resampling~\cite{zhang2024generative}. 
% Thus, QuDT is a direct distance-matching, measurement-assisted generalization of QCBM to ensemble resampling, whereas QuGAN learns the target ensemble through adversarial discrimination.
Therefore, both QuDT and QuGAN can be regarded as one-step generative models.

QuDT and QuGAN both follow the general framework of measurement-assisted evolution, in which the expressivity is ultimately limited by the number of ancilla qubits $N_B$; meanwhile, directly adding more ancilla qubits would significantly increase the complexity of the joint unitary. A natural way to overcome this limitation is to incorporate ancilla mid-circuit measurements and resets, allowing the ancilla qubits to be reused. This leads to the multistep measurement-assisted architecture known as the monitored circuit~\cite{zhang2024generative,zhang2025holographic,zhang2025scaling}. As illustrated in Fig.~\ref{fig:general_sequential}b, the monitored circuit can be considered as a special realization of the general measurement-assisted framework, but built from a sequence of measurement-assisted circuit blocks. By repeatedly measuring and resetting a small ancilla register, the monitored circuit reduces the required physical resources while still retaining sufficient trajectory support and expressive power for generative learning. Its multistep structure decomposes a complex global transport process into a sequence of simpler, more trainable transformations.

{\em QuDDPM.---} QuDDPM adopts the multistep sequential circuit to enable generative learning~\cite{zhang2024generative}. In the forward process, it iteratively applies a random circuit with controllable scrambling power to transform the target state ensemble towards the Haar-random state ensemble. The backward denoising process uses parameterized circuits interleaved with ancilla mid-circuit measurements to implement a reverse mapping from random input states to the target ensemble, with training guided by the forward diffusion process, as sketched in Fig.~\ref{fig:general_sequential}a. This construction naturally admits a stepwise training. Moreover, QuDDPM recasts ensemble generation as learning a multistep transport process rather than a single global map, thereby mitigating training difficulty while retaining the resource efficiency enabled by ancilla reuse.

These one-step and sequential models use different generative resources, including input-state variation and measurement-induced randomness. Comparing them therefore requires task-level metrics that go beyond target-ensemble matching.

\section{Performance Metrics for Ensemble Resampling}
\label{sec:metrics}

Previous studies of quantum generative learning, including QuDDPM~\cite{zhang2024generative}, typically evaluate performance through target-ensemble matching, namely by measuring the distance between the generated ensemble and the target ensemble. While this accuracy-based criterion is essential, it does not by itself reveal how the model reproduces the target ensemble. High target-matching performance may arise because the input ensemble already contains substantial target structure, or because the trained dynamics and measurement trajectories generate the required variation from limited inputs. In measurement-assisted quantum models, output diversity can therefore originate from different sources, including input-state variation and randomness from measurement outcomes. Motivated by this distinction and inspired by evaluation challenges in classical generative modeling~\cite{heusel2017gans,binkowski2018demystifying,zhu2017toward,yang2019diversity,mao2019mode,chen2016infogan}, we introduce three complementary metrics for state-ensemble resampling: accuracy, generative power, and input sensitivity.

{\em Accuracy.}--- {\em Accuracy} quantifies how well the generated state ensemble approximates the target one.     Given a target ensemble $\calE_{\rm tar}$, we define the accuracy $\mathcal{A}$ of the generated state ensemble $\calE_{\rm gen}$  as
\be
    \mathcal{A} \coloneqq 1-\calD\left(\calE_{\rm gen}, \calE_{\rm tar}\right)\in [0,1],
\ee
where $\calD$ is a desired distance measure bounded within $[0,1]$. Ideally, the distance measure $\calD$ needs to satisfy the axioms of a metric, while one may adopt more general functions in practice.
We sketch the idea behind the definition in Fig.~\ref{fig:overall_scheme}a. High accuracy indicates that the unconditional generated ensemble
$\calE_{\rm gen} \simeq \calE_{\rm tar}$ is close to the target ensemble, while low accuracy indicates a mismatch.

Various measures of similarity between state ensembles~\cite{zhang2024generative, yao2026hierarchy} can be adopted here. In this work, we mainly focus on the 2-Wasserstein distance $W_2(\calE_1, \calE_2)$ defined with trace-distance ground metric, introduced in Ref.~\cite{zhang2024generative} as
\be
    W_2(\calE_1, \calE_2) \coloneqq \left(
                                    \inf_{\pi \in \Pi(\mathcal{E}_1,\mathcal{E}_2)}
                                    \mathbb{E}_{(\ket{\psi},\ket{\phi})\sim \pi}\!\left[T(\rho_\psi,\rho_\phi)^2\right]
                                    \right)^{1/2}.
\ee
Here $\rho_\psi=\ketbra{\psi}{\psi}$, $\rho_\phi=\ketbra{\phi}{\phi}$, $T(\rho,\sigma)=\frac{1}{2}\norm{\rho-\sigma}_1$ denotes the trace distance, and $\Pi(\mathcal{E}_1,\mathcal{E}_2)$ denotes the set of all couplings whose marginals are the two ensembles $\mathcal{E}_1$ and $\mathcal{E}_2$. As the Wasserstein distance provides full discriminative power between two state ensembles~\cite{yao2026hierarchy}, it enables the characterization of the accuracy of generated samples. 
% \BZ{**Do you check whether Yao et al. give rigorous proof of full discriminative power? If not, we may want to be conservative.} \RM{**they prove that MMD-k is a k-th moment test, and state that '$W_2(\calE_1, \calE_2) = 0 {\quad \rm iff \quad} \calE_1=\calE_2$'.}

For a typical generative task, the accuracy of a generative model depends on the expressivity of the circuit ansatz, the choice of input ensemble, and the trainability of the model. It will be challenging for models suffering from training difficulty to produce a generated state ensemble with high accuracy.

% \paragraph*{Accuracy.} We begin with \textbf{accuracy}, which quantifies how well the generated ensemble approximates the target distribution. In each shot of the inference procedure, an input state $\ket{\phi_k}$ is sampled from the Haar measure or from a simple reference ensemble, depending on the task. It then evolved through the measurement-assisted circuit, yielding a measurement outcome $\bmz$ and the associated post-measurement state $\ket{\psi_{k\mid \bmz}}$. Defining the composite label $\mu \equiv (k, \bmz)$ and writing $\ket{\psi_\mu} \equiv \ket{\psi_{k\mid \bmz}}$, repeated Monte Carlo sampling over 
% $N$ shots yields the finite sampled set 
% \begin{equation}
%     {\mathcal{S}}_\text{gen}=\bigl\{\ket{\psi_{\mu^{(s)}}} \bigr\}_{s=1}^N, \qquad \mu^{(s)} \equiv (k^{(s)}, \bmz^{(s)}),
% \end{equation}
% which serves as a finite-sample representation of the generated distribution $\mathcal{E}_\text{gen}$. We then quantify its discrepancy from the target ensemble by estimating the approximate 2-Wasserstein distance $W_2(\mathcal{S}_\text{gen}, \mathcal{E}_\text{target})$, using the trace distance as the ground distance metric~\cite{villani2009Optimal, cuturi2013Sinkhorn}. This distance is bounded between 0 and 1 and vanishes if and only if the two distributions coincide. It therefore provides a direct measure of how accurately the model reproduces the target ensemble under the natural inference dynamics of the monitored circuit. \RM{RM: should I add a formal mathematical formula for Wasserstein distance here?}

{\em Generative power.---} High accuracy can be achieved by a trivial model if the input ensemble already closely resembles the target ensemble. We therefore introduce {\em generative power} to quantify how well the model can reproduce the target ensemble when the input resource is restricted. To quantify the intrinsic capability of a model to generate an approximate target state ensemble without entirely relying on the input state resources, we define the {\it generative power} $\mathcal{G}$ as the approximation of the generated state ensemble using the smallest input resource of a single {\em fixed} input state $\ket{\phi_0}$ to the target ensemble
\be
    \mathcal{G}\coloneqq 1-\mathbb{E}_{\ket{\phi_0}}\calD(\calE_{{\rm gen}}(\ket{\phi_0}), \calE_{\rm tar})\in [0,1],
    \label{def:generative_power}
\ee
where $\calE_{{\rm gen}}(\ket{\phi_0})$ is the generated state ensemble from the model with input $\ket{\phi_0}$. $\E_{\ket{\phi_0}}$ denotes the average over the input state ensemble to represent the performance with typical inputs.
As sketched in Fig.~\ref{fig:overall_scheme}a, if the model can generate the entire target ensemble from a single input state, corresponding to $\calG = 1$, we consider it to have the maximum generative power. Intuitively, the generative power of a model is constrained by the total number of ancilla measurements --- $K$ measurements only allow the generation of at most $2^K$ states from a single input. 
% When the {\em generative power} is unity, $\mathcal{G}=1$, it means that the model takes a single input state and produces the entire target ensemble.
%Note that the definition of generative power here is also closely related to the generalization of neural networks~\cite{}. 

{\em Input sensitivity.---} To probe the dependence of model output on the input states, we introduce {\em input sensitivity}. For measurement-assisted quantum generative models, we define the {\em input sensitivity} $\mathcal{S}$ as 
\be
    \mathcal{S}\coloneqq 1-\E_\bmz \overline{F}\left(\calE_{{\rm gen}|\bmz}, \calE_{{\rm gen}|\bmz}\right) \in \left[0,1-\frac{1}{d_A}\right],
    \label{eq:forgetfulness_def}
\ee
where $\calE_{{\rm gen}|\bmz}$ is the generated ensemble conditioned on measurement outcome $\bmz$ introduced in conditional transport. Here, $\overline{F}(\calE_1, \calE_2) \coloneqq \E_{\ket{\psi}\sim \calE_1, \ket{\phi} \sim \calE_2} |\braket{\psi|\phi}|^2$ is the average fidelity between two state ensembles $\calE_1, \calE_2$, and we take the ensemble average over measurement outcomes $\bmz$ to characterize the typical performance. One can also show that $1-\mathcal{S}$ is equal to the average purity of the average state of the ensemble $\calE_{{\rm gen}|\bmz}$. 

As schematically shown in Fig.~\ref{fig:overall_scheme}a, {\em input sensitivity} quantifies the diversity of generated samples originating {\em only} from different inputs. 
If the state samples in this conditional ensemble $\calE_{{\rm gen}|\bmz}$ are approximately identical with high pairwise fidelity, then input sensitivity $\mathcal{S}\simeq 0$, indicating that input variation no longer contributes appreciably to the diversity of generated outputs. As we will show in Section~\ref{sec:concentration}, this situation naturally arises in deep sequential monitored circuits: repeated conditioning on measurement outcomes can make the final state increasingly determined by the measurement trajectory and circuit dynamics, rather than by the initial input state. Importantly, low input sensitivity does not imply low diversity of the unconditional generated ensemble; rather, it indicates that this diversity is supplied primarily by measurement trajectories instead of input-state variation.

% \RM{The three metrics give the qualitative expectations summarized in Fig.~\ref{fig:overall_scheme}(d). One-step models such as QuDT and QuGAN have a limited measurement trajectory space for fixed ancilla size, so their accuracy and generative power can be constrained, while their conditional output ensembles can still retain substantial input sensitivity. Sequential monitored models, such as QuDDPM, reuse the ancilla through repeated measurements and resets, thereby expanding the measurement trajectory space and potentially improving accuracy and generative power. The same repeated conditioning, however, can shift the dominant source of output diversity from input-state variation toward measurement trajectories, leading to reduced input sensitivity.}

%At the end of this section, we emphasize that the above metrics are currently target-dependent and should be understood with respect to a specified ensemble-resampling task from an input ensemble $\calE_0$ to a target ensemble $\calE_{\rm tar}$. 
At the end of this section, we emphasize that the above metrics are target-dependent and should be understood with respect to a specified ensemble-resampling task from an input ensemble $\calE_0$ to a target ensemble $\calE_{\rm tar}$. This target dependence is not unusual: in classical machine learning, model performance is also evaluated relative to prescribed benchmark datasets and tasks, such as MNIST, CIFAR-10, and ImageNet. In the same spirit, the present metrics are intended to characterize the performance of a quantum generative model on a specified target ensemble, rather than to define a task-independent property of the circuit ansatz alone.

\section{Concentration phenomenon in generative quantum machine learning}
\label{sec:concentration}

We now identify a concentration phenomenon in sequential measurement-assisted dynamics, which is the central mechanism underlying the loss of input sensitivity anticipated above. Following the definition of Eq.~\eqref{eq:forgetfulness_def}, we focus on conditional transport in measurement-assisted models. Intuitively, if two nearly orthogonal input states, $|\braket{\phi_1|\phi_2}|^2 \ll 1$, are mapped to nearly identical conditional output states, $|\braket{\psi_{\bmz|1}|\psi_{\bmz|2}}|^2 \simeq 1$, then the conditional transport has low input sensitivity and exhibits concentration. In the following, we analytically study how this concentration develops with system dimension and temporal depth. The analytical results below are derived for Haar-random monitored circuits and should be interpreted as a typical-dynamics benchmark rather than a theorem for trained variational circuits. We then compare this benchmark with optimized QuDDPM circuits obtained from learning tasks and examine whether the same qualitative concentration behavior is observed in the trained dynamics.

\subsection{One-step measurement-assisted model}
\label{subsec:onestep_concentration}

We begin with the one-step measurement-assisted generative model, consisting of a single global unitary followed by measurement on ancilla qubits, as realized by QuDT and QuGAN. With $\calS_0 = 1-\overline{F}(\calE_0, \calE_0)$ denoting the average infidelity of the input state ensemble, we have the following result for input sensitivity (see proof in Appendix~\ref{app:onestep_concentration})
\begin{theorem}
For one-step measurement-assisted generative models, the ensemble-averaged input sensitivity over Haar-random unitaries in the large-dimension limit $d_A\gg 1$ is 
\begin{align}
    \E_{\mathrm{Haar}}\mathcal{S}_1
    &=
    \calS_0
    -\frac{1}{d_A}
    \left(1-\frac{1}{d_B}\right)
    \Delta_3(\rho_{0})
    +\calO\left(d_A^{-2}\right),
    \label{eq:one_step_haar_avg_sensitivity}
\end{align}
where $\Delta_3(\rho_{0})\equiv 1-2\,\tr(\rho_{0}^3) + \left(1-\calS_0\right)^2$ with $\rho_0$ to be the average state of input ensemble $\calE_0$.
\end{theorem}
Here $d_A$ and $d_B$ are the Hilbert-space dimensions of the system $A$ and ancilla $B$. Eq.~\eqref{eq:one_step_haar_avg_sensitivity} shows that the Haar-averaged input sensitivity of the conditional generated ensemble only decreases by $\calO(1/d_A)$, with the dependence on ancilla size entering only through the bounded prefactor $1-1/d_B$. Thus, a one-step measurement-assisted model with typical unitaries does not substantially reduce the distinguishability of states within the input ensemble.

\subsection{Sequential monitored circuits}
\label{sec:dynamics}

\begin{figure*}[t]

    \includegraphics[width=\textwidth]{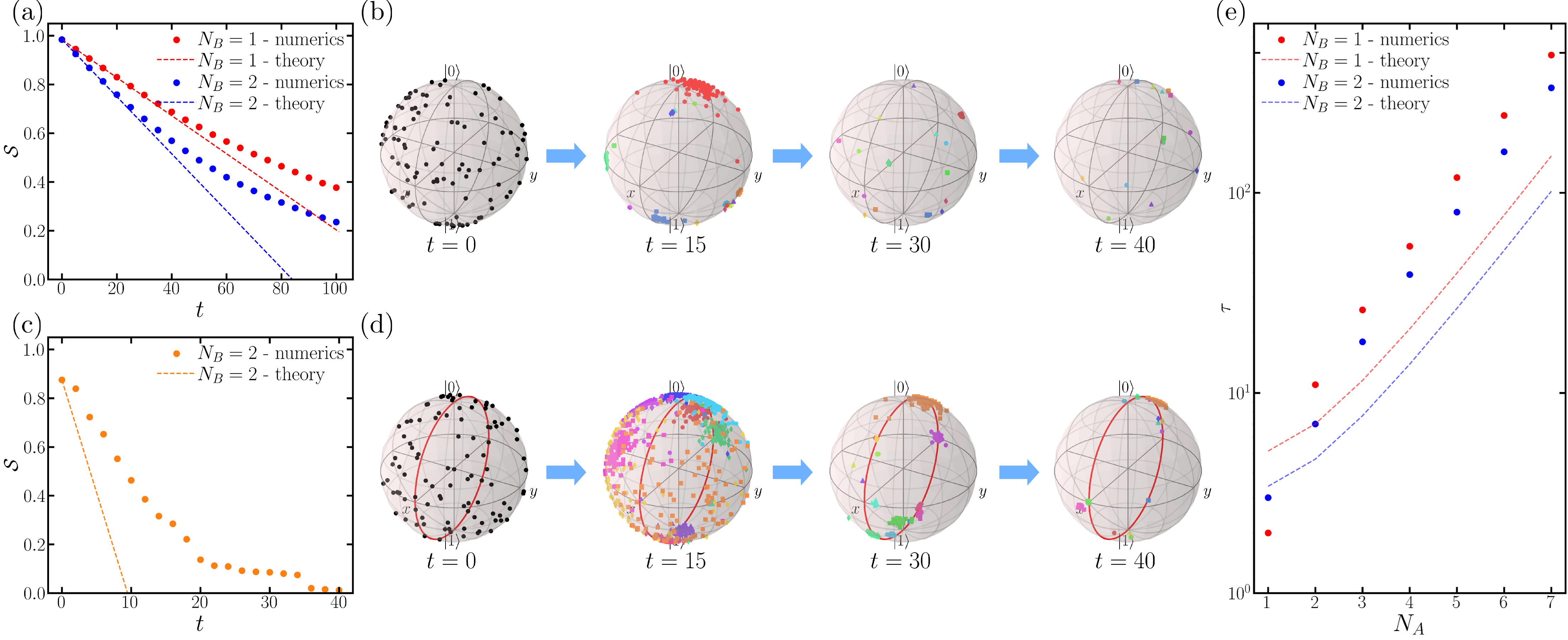}
    % \centering
    \caption{
    (a) Input sensitivity versus temporal depth $t$ in the conditional transport with random quantum circuits of $N_A = 6$ and $N_B = 1, 2$ (red and blue) qubits. Dots and dashed lines represent numerical simulation and theoretical results of Eq.~\eqref{eq:multi_step_haar_avg_sensitivity}.
    (b) Conditional transport with random circuit dynamics on the Bloch sphere for $N_A=1$. Dots with different colors correspond to different trajectories.
    (c) Input sensitivity for a trained QuDDPM circuit, with $N_A=3$ and $N_B=2$. Dots show numerical simulation results; dashed lines show the Haar-random prediction of Eq.~\eqref{eq:multi_step_haar_avg_sensitivity} as a qualitative benchmark rather than a fitted theory for the trained circuit. 
    (d) Conditional transport with trained circuit dynamics on the Bloch sphere for $N_A=1$. The red circle indicates the target ensemble.
    (e) Scaling of the characteristic concentration time $\tau$ versus system size $N_A$ in random circuit dynamics with $N_B = 1, 2$ (red and blue). Here we choose $\epsilon=0.1$. Dots show numerical simulation results and dashed lines represent theoretical results from Eq.~\eqref{eq:concentration_time}.
    }
    \label{fig:concentration_numerics}
\end{figure*}

We now turn to sequential monitored circuits, where repeated measurement-and-reset cycles give rise to the concentration behavior by allowing the loss of input sensitivity to accumulate over steps. For Haar-random monitored circuits, this behavior is captured by the following scaling result (see proof in Appendix~\ref{app:multistep_concentration}):
\begin{theorem}
For multistep measurement-assisted generative models, the ensemble-averaged input sensitivity over Haar-random unitaries in the large-dimension limit $d_A\gg t$ is 
\begin{align} 
    &\E_{\mathrm{Haar}} \mathcal{S}_t %\simeq 1-\frac{1}{2} d_A^{-2t}d_B^{-t} (1+\gamma)(d_A + 1)^{t}(d_A d_B - 1)^t + \nonumber \\
    %&\qquad \frac{1}{2} d_A^{-2t}d_B^{-t} (1-\gamma) (d_A-1)^{t} (d_A d_B + 1)^t  \label{eq:avg_fidelity} \\ 
    = \calS_0 - \frac{t}{d_A} \left(1-\frac{1}{d_B} \right)\Delta_3(\rho_0) + \mathcal{O}\left(d_A^{-2}\right),
    \label{eq:multi_step_haar_avg_sensitivity}
\end{align}
\end{theorem}
We thereby define the characteristic concentration time for $\epsilon$-{\em input sensitivity} by $\E_{\mathrm{Haar}} \mathcal{S}_\tau \le \epsilon$ , which results in
\be
    t \ge \tau \coloneqq \frac{d_A d_B}{(d_B-1)\Delta_3(\rho_0)}(\calS_0-\epsilon) \sim \calO(d_A \calS_0).
    \label{eq:concentration_time}
\ee
For an input state ensemble with $\calS_0 \sim \calO(1)$, i.e., a random state ensemble with exponentially many states, the characteristic time grows exponentially with system size $N_A$. Thus, complete loss of input sensitivity requires exponentially long monitored dynamics for highly diverse input ensembles. Such exponential scaling is also found in the quantum information lifetime of monitored dynamics~\cite{zhang2025holographic}. Beyond this concentration time $t>\tau$, the conditional ensemble enters the overconcentrated regime, in which its states become nearly independent of the initial input state. The derivation does not directly apply to the optimized backward circuit of QuDDPM. Nevertheless, as shown numerically below, optimized QuDDPM circuits exhibit a similar loss of input sensitivity at sufficiently large temporal depth.

% \begin{figure*}[t]
%     \includegraphics[width=\textwidth]{fig/input_sensitivity_plots1.png}
%     % \centering
%     \caption{
%     (a) Average fidelity versus temporal step $t$ in conditional transport in a system of $N_A = 6$ and $N_B = 1, 2$ (red and blue) qubits. (b) Concentration time $\tau$ versus number of system size $N_A$ and different ancilla size $N_B$. Here we choose $\epsilon=0.1$. In both panels, we plot numerical simulation results (circles) and theoretical results (dashed lines) of (a) Eq.~\eqref{eq:multi_step_haar_avg_sensitivity} and (b) Eq.~\eqref{eq:concentration_time}
%     % Concentration in trajectory-resolved conditional transport under typical diffusion. (a) Average purity of the conditional output ensemble versus temporal step $t$ at $N_A=6$ (b) Concentration time $t_c$, defined by $\overline{\gamma}(t_c)=1-\epsilon$ with $\epsilon=0.1$, versus target system size $N_A$. In both panels, simulation and analytical approximation are shown for $N_B=1,2$.
%     \BZ{**[Maybe also show results about optimized circuits if we have it]}\QZ{****I agree, better show both typical and trained}}
%     \label{fig:concentration_numerics}
% \end{figure*}

% \begin{figure}[t]
%     \includegraphics[width=\columnwidth]{fig/input_sensitivity_plots4.png}
%     % \centering
%     \label{fig:concentration_numerics}
% \end{figure}

We numerically verify the dynamics of input sensitivity in conditional transport with random unitaries in Fig.~\ref{fig:concentration_numerics}a for a system of $N_A = 6$ qubits and various ancilla sizes $N_B$. With increasing temporal depth $t$, the input sensitivity $\calS$ drops monotonically, leading to concentration in the asymptotic time limit. Furthermore, our theoretical result of Eq.~\eqref{eq:multi_step_haar_avg_sensitivity} (dashed lines) agrees with numerical results at early times. Although Eq.~\eqref{eq:multi_step_haar_avg_sensitivity} is not a theorem for optimized variational circuits, the trained QuDDPM data in Fig.~\ref{fig:concentration_numerics}c show the same qualitative trend: input sensitivity decreases with temporal depth, but with a slower decay than the Haar-random benchmark. We further show the scaling of concentration time in Fig.~\ref{fig:concentration_numerics}e, where the numerical results (dots) show that the concentration time grows exponentially with system size $N_A$, which aligns with our theoretical prediction of Eq.~\eqref{eq:concentration_time} (dashed lines). Since Eq.~\eqref{eq:concentration_time} gives an asymptotic lower bound on the concentration time, finite-size deviations from the analytical input-sensitivity dynamics do not affect the scaling conclusion.

To provide an intuitive understanding, we consider a $N_A=1$ task of single-qubit learning, which allows us to visualize the conditional state ensemble on the Bloch sphere for both random and trained circuit dynamics in Fig.~\ref{fig:concentration_numerics}b and d, respectively. Starting from a random state ensemble (black dots on the leftmost sphere), the conditional generated state ensemble associated with each measurement trajectory (different colors) converges toward a trajectory-dependent fixed state as the temporal depth increases. At $t=40$, conditional ensembles form distinct input-independent clusters. This indicates that, at late times, the conditional transport becomes insensitive to the initial ensemble and therefore cannot fully convert input-state variation into output diversity.

This concentration mechanism affects not only the dependence on the input state, but also how much of the measurement record remains relevant at late times. Although the conditional output is formally specified by the full measurement trajectory (see Eq.~\eqref{eq:cond_ensemble_def}), its dependence on early outcomes becomes weaker with increasing temporal depth. Equivalently, the evolution may be viewed as an early segment that maps the input ensemble to an unconcentrated intermediate ensemble, followed by a late segment that can drive this intermediate ensemble into concentration. The final output is therefore determined primarily by the late-time trajectory, revealing an emergent temporal locality in the conditional transport dynamics. Consequently, once concentration sets in, further increasing the temporal depth does not necessarily enhance output diversity, because the number of measurement trajectories that effectively contribute to the output diversity saturates.

\section{Metric diagnosis of original QuDDPM and QuDT}
\label{sec:metric_tradeoff_diagnosis}

\begin{figure*}[t]

    \includegraphics[width=0.85\textwidth]{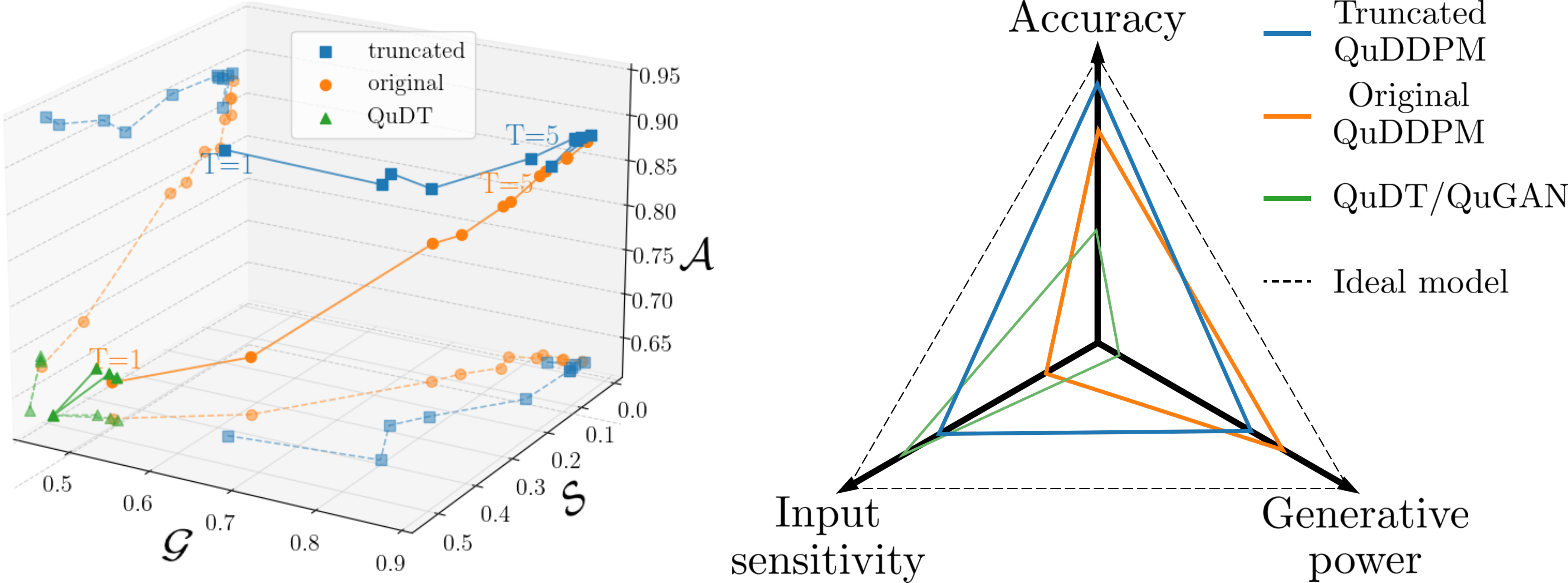}
    % \centering
    \caption{
    Left: Performance benchmark of original QuDDPM and truncated QuDDPM on the logical circular ensemble. The three axes show accuracy $\calA$, generative power $\calG$, and input sensitivity $\calS$. For the original QuDDPM, increasing the temporal depth improves accuracy and generative power by enlarging the number of measurement trajectories, but simultaneously suppresses input sensitivity due to overconcentration. Truncated QuDDPM maintains comparable accuracy over a range of truncation depths while preserving greater input sensitivity, indicating a more favorable balance among the three metrics. The dashed curves are projections onto the $\calS-\calG$ and $\calA-\calS$ planes. Green dots represent the performance of QuDT for reference. Right: schematic of the trade-off between three metrics for different models.
    }
    \label{fig:truncated_full_comparison}
\end{figure*}

Having established concentration as a mechanism that suppresses input sensitivity in monitored circuits, we now examine how this mechanism shapes the performance of trained generative models. We show that the original QuDDPM exhibits a performance-sensitivity trade-off, in which improvements in accuracy and generative power are accompanied by reduced input sensitivity. We investigate the performance of all three metrics proposed in Sec.~\ref{sec:metrics} through the learning of the target state ensemble $\calE_\text{tar} = \{p_\alpha, \ket{\psi(\alpha)}\}_\alpha$ as a representative example, where $\ket{\psi(\alpha)}$ is 
\begin{equation}
    \ket{\psi(\alpha)} = e^{-i\alpha X_L/2}\ket{0_L} = \cos\!\left(\frac{\alpha}{2}\right)\ket{0_L} - i\sin\!\left(\frac{\alpha}{2}\right)\ket{1_L}.
    \label{eq:logical_circular}
\end{equation}
Here $\ket{0_L} = \ket{0}^{\otimes n}, \ket{1_L} = \ket{1}^{\otimes n}$ are logical basis states of the $n$-qubit repetition code, and $X_L = X^{\otimes n}$ is the logical Pauli-$X$. For uniformly distributed $\alpha \sim \mathbb{U}[0, 2\pi]$, the target state ensemble forms a circle in the logical space spanned by $\ket{0_L}, \ket{1_L}$. We thereby refer to it as the logical circular-$X$ ensemble, which generalizes the single-qubit circular state ensemble on the Bloch sphere ~\cite{zhang2024generative, kwun2025mixed} to a multi-qubit logical space. This ensemble serves as a proper benchmark for simultaneously evaluating the performance of a quantum generative learning model across three metrics. For the original QuDDPM, we examine the dependence of its performance on the temporal depth $T$, with properly chosen scrambling strengths for each forward unitary to ensure a smooth interpolation between the target and random state ensembles.

In Fig.~\ref{fig:truncated_full_comparison} left panel, we compare the performance of accuracy $\calA$, generative power $\calG$, and input sensitivity $\calS$ for the original QuDDPMs (orange) and QuDT (green) for generative learning of the 3-qubit logical circular ensemble. For the original QuDDPM, increasing the temporal depth $T$ improves the resampling accuracy, as reflected by the upward trend along the $\calA$ axis. Meanwhile, the number of measurement trajectories grows exponentially with depth, scaling as $\sim 2^{N_B T}$, thereby enhancing the generative power $\calG$. For sufficiently large $T$, the model can therefore generate a diverse set of states from a single input, with output diversity supplied primarily by stochastic mid-circuit measurement outcomes.

However, this gain comes at the cost of input sensitivity: At large $T$, different input states are driven toward nearly identical states given the same measurement trajectory, leading to the suppression of input sensitivity $\calS$ shown in Fig.~\ref{fig:truncated_full_comparison}. This behavior is consistent with the concentration mechanism captured by Eq.~\eqref{eq:multi_step_haar_avg_sensitivity}. Taken together, these results reveal a depth-dependent trade-off in the original QuDDPM: increasing the temporal depth enhances accuracy and generative power through measurement trajectories, but progressively reduces dependence on input-state variation.

% \RM{We next show the performance of QuDT (green), which only involves a single unitary circuit. For a fair comparison, we match its number of trainable parameters to that of the original QuDDPM at each denoising depth $T$. \BZ{**a better description is needed here about circuit setup and parameter count chosen.} QuDT remains limited in generative power and resampling accuracy though preserving high input sensitivity. The weak dependence on parameter count indicates that the main bottleneck is the one-step circuit structure: the lower $\calA$ and $\calG$ reflect limited expressivity and the absence of multistep measurement-induced entanglement, whereas the marginal suppression of $\calS$ is consistent with Eq.~\eqref{eq:one_step_haar_avg_sensitivity}.}

We next compare the results with QuDT (green), a one-step model implemented by a single trainable unitary circuit with ancilla measurement. To isolate the role of temporal depth from the total variational budget, for each denoising depth $T$ of QuDDPM we choose the QuDT circuit such that its number of trainable parameters matches that of the corresponding depth-$T$ QuDDPM ansatz. Thus, varying $T$ for the QuDT only changes the parameter budget of a single-step circuit, rather than introducing a multistep denoising structure. Even under this parameter matching, QuDT remains limited in generative power and resampling accuracy, though it preserves high input sensitivity. The weak dependence on parameter count indicates that the main bottleneck is the one-step circuit structure: the lower $\calA$ and $\calG$ reflect limited expressivity and the absence of a multistep measurement-trajectory space, whereas the marginal suppression of $\calS$ is consistent with Eq.~\eqref{eq:one_step_haar_avg_sensitivity}.

We schematically summarize the trade-off between accuracy, input sensitivity and generative power for various models discussed in Fig.~\ref{fig:truncated_full_comparison} right panel.

% \begin{figure}[b]
%     \centering
%     \includegraphics[width=\columnwidth]{Fig4_tradeoff_3D.png}
%   \caption{
% Performance benchmark of original QuDDPM and truncated QuDDPM on the logical circular ensemble.
% The three axes show accuracy $\calA$, generative power $\calG$, and input sensitivity $\calS$.
% For the original QuDDPM, increasing the temporal depth improves accuracy and generative power by enlarging the number of measurement trajectories, but simultaneously suppresses input sensitivity due to overconcentration.
% Truncated QuDDPM maintains comparable accuracy over a range of truncation depths while preserving greater input sensitivity, indicating a more favorable balance among the three metrics. The dashed curves are projections onto the $\calS-\calG$ and $\calA-\calS$ planes. Green dots represent the performance of QuDT for reference.
% }
% \label{fig:truncated_full_comparison}
% \end{figure}

\section{Mitigating Concentration: Truncated QuDDPM}
\label{sec:truncate_alt}

The overconcentration observed above suggests that the temporal depth of the original QuDDPM should not be increased indefinitely. Increasing the backward temporal depth induces stronger concentration in the conditional generated ensemble, thereby reducing the input-state variation available as a generative resource. At the same time, the original QuDDPM begins the backward process from Haar-random states, whose volume-law entanglement can increase training difficulty through barren-plateau effects~\cite{marrero2021entanglement,cao2025mitigating}. We therefore propose {\em truncated QuDDPM}, which retains the stepwise diffusion training strategy of the original QuDDPM while restricting the temporal depth.

\subsection{Truncated QuDDPM}

\begin{figure}[t]
    \includegraphics[width=\columnwidth]{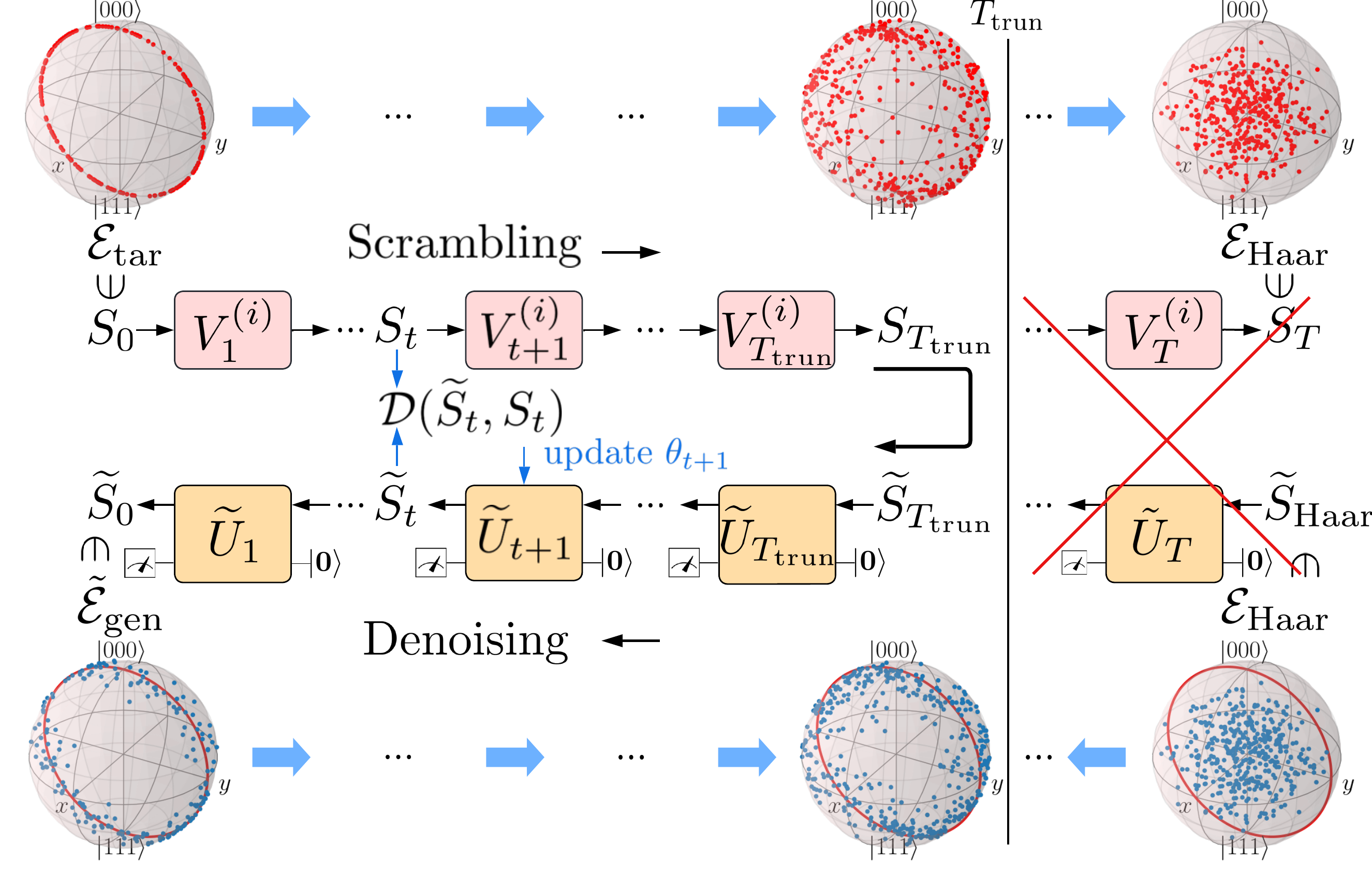}
    % \centering
\caption{
Schematic of truncated QuDDPM.
The original QuDDPM forward process scrambles the target ensemble toward a Haar-like ensemble and then learns a full reverse monitored process. Truncated QuDDPM stops the forward process at an intermediate step $T_{\rm trun}$ and learns only the corresponding shortened reverse process. This reduces the temporal depth responsible for overconcentration while avoiding initialization from highly scrambled states.
}
    \label{fig:truncated_QuDDPM}
\end{figure}

To construct the truncated protocol, we stop the forward diffusion process at an intermediate step $T_{\rm trun}<T$, as illustrated by the vertical line in Fig.~\ref{fig:truncated_QuDDPM}. The resulting partially scrambled ensemble $\tilde S_{T_{\rm trun}}$ serves as the input ensemble for the backward denoising process, replacing the Haar-random input ensemble used in the original QuDDPM. The backward process then learns only the shortened reverse trajectory from $\tilde S_{T_{\rm trun}}$ to the target ensemble. In practice, $\tilde S_{T_{\rm trun}}$ can be prepared by applying the different scrambling unitaries $V_1,\ldots,V_{T_{\rm trun}}$ to the target sample set, without requiring prior knowledge of the full target distribution beyond the available training samples.

The training of each denoising block follows the same layerwise strategy as in the original QuDDPM, but only for the retained temporal depth. After training, the retained reverse process maps samples from $\tilde S_{T_{\rm trun}}$ to states that form the unconditional generated ensemble $\widetilde{\calE}_{\rm gen}$, as shown in the bottom panel of Fig.~\ref{fig:truncated_QuDDPM}. The training procedure is summarized in Algorithm~\ref{alg:truncated_quddpm}. Thus, truncation has two distinct effects: by avoiding initialization from highly entangled Haar-random states, it can reduce training difficulty associated with entanglement-induced barren-plateau effects~\cite{marrero2021entanglement,cao2025mitigating}; by shortening the monitored trajectory, it suppresses the excessive loss of input sensitivity caused by repeated measurement-induced conditioning. The first effect can improve accuracy and generative power, whereas the second effect is independently captured by the input-sensitivity metric.

% Compared to full QuDDPM, the key modification of truncated QuDDPM is that both the forward diffusion and reverse denoising processes are restricted to a retained depth $T_\text{trun}<T$.
% Consequently, rather than from Haar-random states, the reverse process is initialized from the partially scrambled target ensemble $\calS_{T_\text{trun}}$ obtained after $T_\text{trun}$ forward scrambling steps, where the forward scrambling process applies independent random unitaries $V_t^{(i)}$ to each sample with controlled scrambling strength. Training then proceeds in a layerwise backward manner: for each $t=T_\text{trun}-1,T_\text{trun}-2,\ldots,0$, we construct the target noisy ensemble $\calS_t$ by applying the forward scrambling process to the target ensemble up to step $t$. Starting again from $\calS_{T_\text{trun}}$, we apply the retained denoise maps, implemented by the monitored circuits with $\tilde U_{T_\text{trun}},\tilde U_{T_\text{trun}-1},\ldots,\tilde U_{t+1}$, to obtain a reconstructed ensemble $\tilde{\calS}_t$. The parameters of only the next block, $\tilde U_{t+1}(\theta_{t+1})$, are then optimized by minimizing the discrepancy $\cD(\tilde{\calS}_k,\calS_k)$, while the deeper retained blocks are kept fixed. Repeating this procedure for decreasing $t$ progressively trains the truncated reverse process from depth $T_\text{trun}$ back to the target ensemble. 

\begin{algorithm}[H]
\caption{Training of truncated QuDDPM}
\label{alg:truncated_quddpm}
\begin{algorithmic}[1]
    \Statex \textbf{Input:} Target sample set $S_{\rm tar}$ sampled from $\mathcal{E}_{\rm tar}$; retained temporal depth $T_{\rm trun}$
    \Statex \textbf{Output:} Trained denoising maps $\{\widetilde U_j(\theta_j^*)\}_{j=1}^{T_{\rm trun}}$, where $\widetilde U_j$ maps $S_j$ to $S_{j-1}$

    \State Set $S_0 \leftarrow S_{\rm tar}$.
    \For{$t=0,1,\ldots,T_{\rm trun}-1$}
        \State Apply one scrambling step to $S_t$ to obtain $S_{t+1}$.
    \EndFor

    \For{$t=T_{\rm trun}-1,T_{\rm trun}-2,\ldots,0$}
        \State Set $\widetilde S_{T_{\rm trun}} \leftarrow S_{T_{\rm trun}}$.

        \For{$j=T_{\rm trun},T_{\rm trun}-1,\ldots,t+2$}
            \State Apply the trained denoising map $\widetilde U_j(\theta_j^*)$ to $\widetilde S_j$ to obtain $\widetilde S_{j-1}$.
        \EndFor

        \State Apply the trainable denoising map $\widetilde U_{t+1}(\theta_{t+1})$ to $\widetilde S_{t+1}$ to obtain $\widetilde S_t$.
        \State Compute the loss $\mathcal{D}(\widetilde S_t,S_t)$.
        \State Update only $\theta_{t+1}$ to minimize $\mathcal{D}(\widetilde S_t,S_t)$, while keeping $\{\theta_j^*\}_{j=t+2}^{T_{\rm trun}}$ fixed.
        \State Set the optimized parameter as $\theta_{t+1}^*$.
    \EndFor
\end{algorithmic}
\end{algorithm}

% Algorithm~\ref{alg:truncated_quddpm} trains the retained denoising blocks in a layerwise backward manner. Starting from the partially scrambled ensemble $S_K$, at each iteration we target an earlier noisy ensemble $S_k$ and optimize only the monitored circuit $\tilde U_{k+1}(\theta_{k+1})$, while keeping the deeper retained blocks fixed. The loss $\cD(\tilde{\cE}_k,S_k)$ measures how well the retained reverse process reconstructs the ensemble at level $k$. In this way, the truncated protocol learns only the last $K$ reverse steps, rather than the full reverse trajectory from Haar-random states.

\subsection{Performance of truncated QuDDPM}
\label{subsec:truncate_alt_comparison}

We now evaluate the performance of the truncated QuDDPM and compare it with the original QuDDPM. We consider the same logical circular ensemble as Eq.~\eqref{eq:logical_circular}. For the original QuDDPM, whose depth dependence was examined in Sec.~\ref{sec:metric_tradeoff_diagnosis}, the scrambling schedule is chosen separately for each total depth $T$ to ensure a smooth interpolation between the target and random state ensembles. For the truncated QuDDPM, we keep the scrambling strength fixed across different choices of $T_{\rm trun}$, so that changes in performance reflect the effect of truncation depth.

Figure~\ref{fig:truncated_full_comparison} shows the accuracy $\calA$, generative power $\calG$, and input sensitivity $\calS$ of the truncated QuDDPM. Across different choices of truncation depth $T_{\rm trun}$, we observe consistently high accuracy for the truncated model, with $\calA \simeq 0.9$. At smaller temporal depths, the backward process starts from an ensemble closer to the target, making the reverse learning less demanding than learning a full map from Haar-random states. This closer starting point partly accounts for the high accuracy of the truncated model, in contrast to the original QuDDPM at small $T$, where the sharper interpolation toward the Haar-random ensemble leads to poor generation performance, with accuracy $\calA \le 0.7$ for $T\le 2$. However, high accuracy alone would not distinguish genuine resampling from a trivial reliance on the partially scrambled input ensemble. The generative-power metric addresses this point: the truncated model also achieves high $\calG$, indicating that even when inference is restricted to a single fixed input state, the trained monitored dynamics can still reconstruct the target ensemble. Thus, the truncated model does not simply inherit the target structure from its input ensemble, but learns to use measurement trajectories to resample the ensemble. This advantage should be distinguished from the improvement in input sensitivity: avoiding excessive temporal depth does not by itself guarantee higher accuracy or generative power, but it preserves input-dependent variation that is invisible to accuracy alone.

As $T_{\rm trun}$ increases, the truncated model approaches the original QuDDPM and gains access to a larger measurement-trajectory space. This generally improves its generative power $\calG$, since the number of possible trajectories grows with retained depth. For a limited $T_{\rm trun}$, however, the truncated QuDDPM already exhibits stronger generative power than the original QuDDPM, a property that benefits from the partially scrambled starting point. At the same time, increasing $T_{\rm trun}$ also reduces input sensitivity for the truncated model, because longer monitored trajectories induce stronger concentration. Therefore, the truncated QuDDPM still exhibits a trade-off between $\calG$ and $\calS$, but this trade-off can be tuned without strongly degrading accuracy. This suggests an intermediate operating regime in temporal depth for the truncated model, where it can effectively leverage input-state variation while mitigating overconcentration and maintaining good generative power and high accuracy. We summarize the above findings qualitatively in Fig.~\ref{fig:truncated_full_comparison} right panel. Indeed, the main benefit of truncation is not simply improved target matching, but a more favorable operating point across the three complementary metrics.

% As the depth increases from left to right, both models move toward higher concentration. For {\em original} QuDDPM, the short-depth regime corresponds to a highly radical interpolation schedule, making the reverse process difficult to learn and yielding poor accuracy and generative power in the shallow regime. Although these metrics improve with increasing depth, the model tends to rely on strong concentration, which is detrimental to generative diversity. By contrast, for {\em truncated} QuDDPM, the backward circuit at short depth starts from an input ensemble that is already close to the target manifold, since only the retained portion of the scrambling trajectory must be inverted. This gives the truncated model an inherent advantage in accuracy at low depth. As the depth increases, the truncated model further improves its generative power while maintaining strong accuracy throughout. Overall, truncation shifts the accessible performance frontier: it achieves better agreement with the target ensemble and better generative power over most of the scan, without requiring the strong concentration seen in original QuDDPM.

\section{Discussion and Outlook}
\label{sec:discussion}

In this work, we identified a concentration mechanism in measurement-assisted quantum generative learning for state-ensemble resampling. Through input sensitivity, we showed that repeated measurement-induced conditioning can suppress the dependence of conditional outputs on the input ensemble, a feature not captured by target-ensemble matching alone. While increasing temporal depth can improve accuracy and generative power through measurement trajectories, repeated conditioning can reduce input sensitivity. We used this diagnosis to introduce truncated QuDDPM as an example operating regime: by reducing temporal depth, the model preserves stronger input sensitivity while maintaining comparable accuracy and generative power. More broadly, these results highlight temporal depth as a key design parameter for balancing generative performance against overconcentration.

Several future directions naturally follow from this work. On the application side, one can develop more systematic strategies for identifying the optimal truncation point, including adaptive truncation protocols and feedback-controlled designs that adjust the generative process based on intermediate measurement outcomes. Beyond quantum state ensemble resampling, it would be interesting to use the proposed metrics to quantify the performance of quantum models in other generative learning tasks, e.g., classical data such as images~\cite{cacioppo2023quantum, ma2025quantum}. Finally, the proposed metrics could be complemented by standardized test ensembles, allowing one to define more standardized performance benchmarks based on worst-case, best-case, or average-case behavior across a prescribed family of target distributions.

\section*{Acknowledgments}

R.M., B.Z., and Q.Z. acknowledge support from NSF (CCF-2240641, 2350153, OMA-2326746). Q.Z. also acknowledges support from ONR MURI N000142612102, AFOSR MURI FA9550-24-1-0349, and DOE ARPA-E Grant No. DE-AR0002067.

\section*{Data Availability}

% APS expects a data-availability statement for data needed to verify or replicate the work.
The data and code that support the findings of this study are available at \cite{Github}.

\appendix

\begin{widetext}
\section{Derivation of the Monitored-Circuit Concentration}
\label{app:concentration_derivation}

In this appendix, we derive the Haar-averaged conditional-ensemble purity that appears in the input-sensitivity analysis. The derivation follows the notation of the main text. Let the input ensemble be
\be
    \calE_0=\{p_k,\ket{\psi_k}\}_k, \qquad \rho_{0}=\sum_k p_k \state{\psi_k}_{A},
\ee
where $A$ denotes the system register and $B$ denotes the ancilla register. We write $d_A=\dim\calH_A$ and $d_B=\dim\calH_B$. The average fidelity of the input ensemble with itself is
\be
    \gamma \equiv \overline{F}(\calE_0,\calE_0) = \tr(\rho_{0}^2). \label{eq:appendix_input_gamma}
\ee
The input sensitivity $\calS$ is defined as
\begin{align}
    \calS &\equiv 1-\E_{\bmz}\overline{F}(\calE_{{\rm gen}|\bmz},\calE_{{\rm gen}|\bmz}) \\
          & \equiv 1 - \overline{\gamma}
\end{align}

\subsection{Conditional state and replica representation}

For a single monitored block, the ancilla is initialized in $\ket{0}_B$, the input state in the system register is $\rho_0$, the joint unitary $U$ acts on $AB$, and the ancilla is measured in the computational basis. Conditioned on outcome $\bmz$, the unnormalized output state on $A$ is
\be
    \tilde{\rho}_{A|\bmz} = \tr_B\!\left[ U(\rho_0\otimes \state{0}_B)U^\dagger (\bI_A\otimes \state{\bmz}_B) \right], \qquad p_{\bmz}=\tr(\tilde{\rho}_{A|\bmz}). \label{eq:appendix_one_step_conditional_state}
\ee
The normalized branch state is $\rho_{A|\bmz}=\tilde{\rho}_{A|\bmz}/p_{\bmz}$. Thus,
\be
    \overline{\gamma} = \sum_{\bmz}p_{\bmz}\tr(\rho_{A|\bmz}^2) = \sum_{\bmz}p_{\bmz}^{-1}\tr(\tilde{\rho}_{A|\bmz}^2). \label{eq:appendix_purity_with_negative_power}
\ee
To handle the inverse probability, we introduce the pseudo-measurement-conditioned purity using the replica trick,
\be
    \overline{\gamma}^{(m)} \equiv \sum_{\bmz}p_{\bmz}^{m}\tr(\tilde{\rho}_{A|\bmz}^2), \label{eq:appendix_replica_limit}
\ee
% We use $\tau_A$ for the swap operator acting on the last two replicas of $A$, and define
% \be
% D_B^{(n)} \equiv \sum_{\bmz}\state{\bmz}_B^{\otimes n}. \label{eq:appendix_D_definition}
% \ee

We now rewrite the two ingredients in Eq.~\eqref{eq:appendix_replica_limit} in replica form. First, the branch probability $p_\bmz$ satisfies
\be
    p_{\bmz}^{m} = \tr\!\left[ U^{\otimes m} \rho_0^{\otimes m}\state{0}_B^{\otimes m} (U^\dagger)^{\otimes m} \bI_A^{\otimes m}\state{\bmz}_B^{\otimes m} \right]. \label{eq:appendix_probability_replica}
\ee
Second, the unnormalized branch purity can be written with the swap trick as
\begin{align}
    \tr(\tilde{\rho}_{A|\bmz}^2) 
    &= \tr\!\left[ \left(\tilde{\rho}_{A|\bmz}\otimes\tilde{\rho}_{A|\bmz}\right)\tau_A \right] \nonumber\\
    &= \tr\!\left[ U^{\otimes 2} \rho_0^{\otimes 2}\state{0}_B^{\otimes 2} (U^\dagger)^{\otimes 2} \tau_A\state{\bmz}_B^{\otimes 2} \right], \label{eq:appendix_purity_replica}
\end{align}
where $\tau_A$ acts only on the two purity replicas of system $A$. Multiplying Eqs.~\eqref{eq:appendix_probability_replica} and~\eqref{eq:appendix_purity_replica} combines the $m$ probability replicas and the two purity replicas into a single $n=m+2$ replica trace:
\begin{align}
    \overline{\gamma}^{(m)} 
    &= \sum_{\bmz}\tr\!\left[ U^{\otimes (m+2)} \rho_0^{\otimes (m+2)}\state{0}_B^{\otimes (m+2)} (U^\dagger)^{\otimes (m+2)} \left(\bI_A^{\otimes m}\otimes\tau_A\right)  \state{\bmz}_B^{\otimes (m+2)}\right]\\ 
    &=\tr\!\left[ U^{\otimes n} \rho_0^{\otimes n}\state{0}_B^{\otimes n} (U^\dagger)^{\otimes n} \left(\bI_A^{\otimes m}\otimes\tau_A\right) D_B^{(n)} \right]. \label{eq:appendix_one_step_replica_purity}
\end{align}
where $D_B^{(n)} \equiv \sum_{\bmz}\state{\bmz}_B^{\otimes n}$.

% For a depth-$t$ monitored circuit, the ancilla is reset to $\ket{0}$ after each measurement. The circuit consists of independent Haar-random unitaries $U_1,\ldots,U_t$ acting on $A_iB_i$. After summing over the full trajectory $\bmz=(\bmz_1,\ldots,\bmz_t)$, the replica network is obtained by iterating Eq.~\eqref{eq:appendix_one_step_replica_purity}. The final boundary condition is $\left(\bI_A^{\otimes m}\otimes\tau_A\right)D_B^{(n)}$, while each earlier measured ancilla contributes another copy of $D_B^{(n)}$. Equivalently, if $\tilde{\rho}_{A_i}^{(n)}$ denotes the unnormalized $n$-replica state after the first $i$ monitored blocks and after summing over the first $i$ outcomes, then
% \begin{align}
% \tilde{\rho}_{A_i}^{(n)} &= \tr_{B_i}\!\left[ U_i^{\otimes n} \left(\tilde{\rho}_{A_{i-1}}^{(n)} \otimes \state{0}_{B_i}^{\otimes n}\right) (U_i^\dagger)^{\otimes n} \bI_{A_i}^{\otimes n}D_{B_i}^{(n)} \right], \label{eq:appendix_replica_recursion}\\
% \tilde{\rho}_{A_0}^{(n)} &= \rho_{A_0}^{\otimes n}.
% \end{align}
% The depth-$t$ replica purity is then
% \be
% \overline{\gamma}_t^{(m)} = \tr\!\left[ \tilde{\rho}_{A_t}^{(n)} \left(\bI_{A_t}^{\otimes m}\otimes\tau_{A_t}\right) \right]. \label{eq:appendix_depth_t_replica_purity}
% \ee
% This form makes clear why each monitored step contributes the same Haar-averaged transfer rule after vectorization.

\subsection{Haar twirl and standard first-shell power counting}

We now average each monitored block over the Haar measure. For the vectorized Hilbert-Schmidt notation $\roundket{M}=\sum_{ab}M_{ab}\ket{a}\ket{b}$,
\be
    \tr(M^\dagger N)=\roundbra{M}\roundket{N}, \qquad \roundket{AMB}=(A\otimes B^T)\roundket{M}, \qquad \roundbra{M}(A\otimes B^T)\roundket{N} = \tr(M^\dagger ANB),
\ee

We denote the corresponding swap final boundary by $ \roundket{\tau}\roundket{D_{B}^{(n)}}$, where $\roundket{\tau}_A \equiv \roundket{\bI_A^{\otimes m}\otimes\tau_A}$, and rewrite Eq.~\eqref{eq:appendix_one_step_replica_purity} as 

\begin{align}
    \E_\text{Haar}\overline{\gamma}^{(m)} &= \E_\text{Haar}\tr\!\left[ U^{\otimes n} \rho_0^{\otimes n}\state{0}_B^{\otimes n} (U^\dagger)^{\otimes n} \left(\bI_A^{\otimes m}\otimes\tau_A\right) D_B^{(n)} \right] \nonumber \\
    &= \E_\text{Haar} \roundbra{\rho_{0}}_A^{\otimes n} \roundbra{0}_B^{\otimes n}\left[U^{\otimes n} \otimes (U^{*})^{\otimes n} \right] \roundket{\tau}_A\roundket{ D_B^{(n)}} \nonumber \\
    &= \roundbra{\rho_{0}}_A^{\otimes n} \roundbra{0}_B^{\otimes n}\E_\text{Haar}\left[U^{\otimes n} \otimes (U^{*})^{\otimes n} \right] \roundket{\tau}_A\roundket{ D_B^{(n)}}
\end{align}

Next, we evaluate the Haar twirling on $n$ replicas by Weingarten calculus,
\begin{align}
    \Phi_n &\equiv \E_{U\sim{\rm Haar}} \left[ U^{\otimes n}\otimes (U^*)^{\otimes n} \right] \nonumber \\
    &= \sum_{\sigma,\pi\in S_n} {\rm Wg}(\sigma^{-1}\pi,d) \roundket{\hat{\sigma}}\roundbra{\hat{\pi}} \nonumber \\
    &= {\rm Wg}(e,d) \sum_{\pi\in S_n} \roundket{\hat{\pi}}\roundbra{\hat{\pi}} + \sum_{\substack{\sigma,\pi\in S_n\\ \sigma\neq\pi}} {\rm Wg}(\sigma^{-1}\pi,d) \roundket{\hat{\sigma}}\roundbra{\hat{\pi}},
    \label{eq:appendix_weingarten_split}
\end{align}
where $\hat{\sigma}$ and $\hat{\pi}$ denote the permutation operators associated with $\sigma,\pi\in S_n$ on the full replica space $AB$ and $d=d_Ad_B$. 

To estimate $\E_\text{Haar} \overline \gamma^{(m)}$, the final purity boundary contains the swap on the last two $A$ replicas. Applying one monitored Haar block to the final boundary $\roundket{\tau}_A\roundket{D_{B}^{(n)}}$ and attaching the initial ancilla $B$-boundary $\roundbra{0}_{B}^{\otimes n}$, we obtain
\begin{align}
    \roundbra{0}_B^{\otimes n}\Phi_{\rm n}\roundket{\tau}_A\roundket{D_{B}^{(n)}} 
    &= d_B{\rm Wg}(e,d) \sum_{\pi\in S_n} d_A^{\#(\pi^{-1}\tau)} \roundket{\hat{\pi}}_{A} \nonumber\\
    &\quad+ d_B \sum_{\substack{\sigma,\pi\in S_n\\ \sigma\neq\pi}} {\rm Wg}(\sigma^{-1}\pi,d) d_A^{\#(\pi^{-1}\tau)} \roundket{\hat{\sigma}}_{A}, \label{eq:appendix_monitored_split_tau}
\end{align}
where we used $\roundbra{\hat{\pi}}\roundket{D_{B}^{(n)}}_B =\tr(\pi^\dagger D_{B}^{(n)})=d_B$ and $\roundbra{\hat{\pi}}\roundket{\tau}_A =d_A^{\#(\pi^{-1}\tau)}$, as well as $\roundbra{0}_B^{\otimes n}\hat{\pi}_{B} =\roundbra{0}^{\otimes n}_B$ for any replica permutation $\pi$, and therefore $\roundbra{0}_{B}^{\otimes n} \roundket{\hat{\pi}}_{AB} = \roundket{\hat{\pi}}_{A}$.

We use the standard large-dimension asymptotic expansion of the unitary Weingarten function,
\be
    {\rm Wg}(\omega,d) = d^{-2n+\#(\omega)} \left[ {\rm Mob}(\omega)+\calO(d^{-2}) \right], \qquad d=d_Ad_B, \label{eq:appendix_wg_asymptotic}
\ee
where $\#(\omega)$ denotes the number of cycles of the permutation $\omega \in S_n$, and ${\rm Mob}(\omega)$ is the corresponding Mobius factor. In particular, ${\rm Mob}(e)=1$ for identity $e$ and ${\rm Mob}(\omega)=-1$ for a transposition $\omega$. The expansion is understood in the fixed-replica-number regime, with $n=m+2$ and fixed $\omega$, followed by the large-dimension limit $d=d_A d_B\to\infty$.

The diagonal contribution is then organized by the number of cycles in $\pi^{-1}\tau$. Denoting $\ell = n-\#(\pi^{-1}\tau)$, one diagonal term has size
\begin{align}
    d_B{\rm Wg}(e,d)d_A^{\#(\pi^{-1}\tau)} 
    &= d_B\left(d^{(-n)} +\calO(d^{-n-2}) \right)d_A^{\#(\pi^{-1}\tau)} \nonumber \\
    &= d_A^{-\ell}d_B^{1-n}\left(1+\calO(d^{-2})\right) , \label{eq:appendix_diagonal_power_count}
\end{align}
The leading orders of the first few diagonal shells are
\begin{table}[h!]
\caption{
Power counting of the diagonal Weingarten contribution $d_B {\rm Wg}(e,d)d_A^{\#(\pi^{-1}\tau)}$, with $\ell=n-\#(\pi^{-1}\tau)$.
}
\label{tab:appendix_diagonal_power_count}
\begin{ruledtabular}
\begin{tabular}{cccc}
Permutation class $\pi$ & \# cycles & $\ell$ & Leading order \\
\hline
$\pi=\tau$
& $n$
& $0$
& $\calO(d_B^{1-n})$ \\

$\pi=\tau s,\; s\in\calT_1$
& $n-1$
& $1$
& $\calO(d_A^{-1}d_B^{1-n})$ \\

$\#(\pi^{-1}\tau)=n-2$
& $n-2$
& $2$
& $\calO(d_A^{-2}d_B^{1-n})$
\end{tabular}
\end{ruledtabular}
\end{table}

Thus, the unique leading term is $\pi=\tau$, and the first subleading shell is $\pi=\tau s$ with $s\in\calT_1$, while higher-order terms are suppressed in the limit $d_A \gg 1$. Keeping the leading shells in the large-$d_A$ expansion gives
\be
    d_B{\rm Wg}(e,d) \sum_{\pi\in S_n} d_A^{\#(\pi^{-1}\tau)} \roundket{\hat{\pi}}_{A}
    =
    d_B^{1-n}\!\left[\roundket{\hat{\tau}}_{A}+\frac{1}{d_A}\sum_{s\in\calT_1}\roundket{\widehat{\tau s}}_{A}+\calO(d_A^{-2})\right].
    \label{eq:appendix_diagonal_standard_shells}
\ee

We then perform the same approximation on the off-diagonal part,
\be
    d_B \sum_{\substack{\sigma,\pi\in S_n\\ \sigma\neq\pi}} {\rm Wg}(\sigma^{-1}\pi,d) d_A^{\#(\pi^{-1}\tau)} \roundket{\hat{\sigma}}_{A} . \label{eq:appendix_offdiagonal_after_B_connection}
\ee
The off-diagonal contribution is organized by both the separation between the two Weingarten permutations and the distance from the final swap boundary. Denoting
\[
r=n-\#(\sigma^{-1}\pi), \qquad \ell=n-\#(\pi^{-1}\tau),
\]
one off-diagonal term has size
\begin{align}
    d_B{\rm Wg}(\sigma^{-1}\pi,d)d_A^{\#(\pi^{-1}\tau)}
    &= d_B d^{-n-r}\left[{\rm Mob}(\sigma^{-1}\pi)+\calO(d^{-2})\right] d_A^{n-\ell} \nonumber \\
    &= \calO\!\left(d_A^{-r-\ell}d_B^{1-n-r}\right). \label{eq:appendix_offdiagonal_power_count}
\end{align}
The leading orders of the first few off-diagonal shells are
\begin{table}[h!]
\caption{
Power counting of the off-diagonal Weingarten contribution
$d_B{\rm Wg}(\sigma^{-1}\pi,d)d_A^{\#(\pi^{-1}\tau)}$, with
$r=n-\#(\sigma^{-1}\pi)$ and $\ell=n-\#(\pi^{-1}\tau)$.
}
\label{tab:appendix_offdiagonal_power_count}
\begin{ruledtabular}
\begin{tabular}{cccc}
Permutation class $(\sigma,\pi)$ & $r$ & $\ell$ & Leading order \\
\hline
$\pi=\tau,\; \sigma=\tau s,\; s\in\calT_1$
& $1$
& $0$
& $\calO(d_A^{-1}d_B^{-n})$ \\

$\pi=\tau s_1,\; \sigma=\pi s_2,\; s_1,s_2\in\calT_1$
& $1$
& $1$
& $\calO(d_A^{-2}d_B^{-n})$ \\

$\pi=\tau,\; n-\#(\sigma^{-1}\tau)=2$
& $2$
& $0$
& $\calO(d_A^{-2}d_B^{-n-1})$
\end{tabular}
\end{ruledtabular}
\end{table}

Thus, the first off-diagonal shell is $\pi=\tau$ and $\sigma=\tau s$ with $s\in\calT_1$, while higher shells are at least order $\calO(d_A^{-2})$. Keeping this shell gives
\be
    d_B \sum_{\substack{\sigma,\pi\in S_n\\ \sigma\neq\pi}} {\rm Wg}(\sigma^{-1}\pi,d) d_A^{\#(\pi^{-1}\tau)} \roundket{\hat{\sigma}}_{A}
    = -\frac{1}{d_A}d_B^{-n}\sum_{s\in\calT_1}\roundket{\widehat{\tau s}}_{A}+\calO(d_A^{-2}d_B^{1-n}),
    \label{eq:appendix_offdiagonal_standard_shell}
\ee
where we used ${\rm Mob}(s)=-1$ for any transposition $s$. Combining the diagonal and off-diagonal contributions through the first subleading order gives the boundary-connected rule
\begin{align}
    \roundbra{0}_{B}^{\otimes n} \Phi_n \roundket{\tau}_A\roundket{D_{B}^{(n)}} 
    &=\left[ q_0\roundket{\tau}_A + q_1 \sum_{s\in\calT_1} \roundket{\widehat{\tau s}}_A + \calO(d_A^{-2}d_B^{-(m+1)}) \right], \label{eq:appendix_standard_first_shell} \\
    \text{where}\quad q_0 &= d_Bd^{-n}d_A^n = d_B^{-(m+1)}, \nonumber\\
    q_1 &= d_B\left[ d^{-n}d_A^{n-1} - d^{-n-1}d_A^n \right] = \frac{1}{d_A}d_B^{-(m+1)} \left(1-\frac{1}{d_B}\right). \nonumber 
\end{align}

\subsection{One-step boundary contraction}
\label{app:onestep_concentration}

We now evaluate the one-step boundary contraction by attaching the input ensemble boundary $\roundbra{\rho_{0}}_A^{\otimes n}\roundbra{0}_B^{\otimes n}$ and final swap boundary $\roundket{\tau}_A\roundket{D_B^{(n)}}$ to the Haar twirling. Thus
\be
    \E_{\rm Haar}\overline{\gamma}^{(m)}_1 = \roundbra{\rho_{0}}_A^{\otimes n} \roundbra{0}_B^{\otimes n} \Phi_n \roundket{\tau_A} \roundket{D_B^{(n)}} . \label{eq:appendix_one_step_twirl_boundary}
\ee
Using Eq.~\eqref{eq:appendix_standard_first_shell} inside the inner product in Eq.~\eqref{eq:appendix_one_step_twirl_boundary} gives
\begin{align}
    \E_{\rm Haar}\overline{\gamma}^{(m)}_1 
    &= q_0 \roundbra{\rho_{0}}_A^{\otimes n} \roundket{\tau}_A + q_1 \sum_{s\in\calT_1} \roundbra{\rho_{0}}_A^{\otimes n} \roundket{\widehat{\tau s}}_A + \calO(d_A^{-2}d_B^{-(m+1)}) . \label{eq:appendix_one_step_inner_product_expanded}
\end{align}
Here the $B$ contraction has already been included in Eq.~\eqref{eq:appendix_standard_first_shell}; only the $A$ inner products remain. For the remaining $A$ contraction with the input ensemble, define
\be
    M_\sigma(\rho_{0}) \equiv \roundbra{\rho_{0}}_A^{\otimes n} \roundket{\hat{\sigma}}_A = \prod_{c\in \mathrm{cycles}(\sigma)} \operatorname{Tr}\!\left(\rho_{0}^{|c|}\right).  \label{eq:appendix_input_moment_definition}
\ee
In particular,
\be
    M_\tau(\rho_{0}) = \tr(\rho_{0}^2) = \gamma, \qquad M_e(\rho_{0}) = 1 . \label{eq:appendix_input_moment_special_cases}
\ee
Therefore Eq.~\eqref{eq:appendix_one_step_inner_product_expanded} becomes
\begin{align}
    \E_{\rm Haar}\overline{\gamma}^{(m)}_1 &= q_0\gamma + q_1 \sum_{s\in\calT_1} M_{\tau s}(\rho_{0}) + \calO(d_A^{-2}d_B^{-(m+1)}) \nonumber\\
    &= d_B^{-(m+1)}\gamma + \frac{1}{d_A}d_B^{-(m+1)} \left(1-\frac{1}{d_B}\right) \sum_{s\in\calT_1} M_{\tau s}(\rho_{0}) + \calO(d_A^{-2}d_B^{-(m+1)}) \nonumber\\
    &= d_B^{-(m+1)}\gamma + \frac{1}{d_A}d_B^{-(m+1)} \left(1-\frac{1}{d_B}\right) \left[ 1 + \sum_{\substack{s\in\calT_1\\ s\neq\tau}} M_{\tau s}(\rho_{0}) \right] + \calO(d_A^{-2}d_B^{-(m+1)}) . \label{eq:appendix_one_step_first_shell}
\end{align}
In the last line we separated the $s=\tau$ term, which gives $M_{\tau^2}(\rho_0)=M_e(\rho_0)=1$. The remaining transpositions $s\neq\tau$ contribute the other first-shell moments at the same order. Their sum can be estimated from the cycle structure of $\tau s$. If $s$ shares one replica index with $\tau$, then $\tau s$ is a three-cycle and $M_{\tau s}(\rho_0)=\tr(\rho_0^3)$; there are $2(n-2)$ such transpositions. If $s$ is disjoint from $\tau$, then $\tau s$ is a product of two disjoint transpositions and $M_{\tau s}(\rho_0)=\tr(\rho_0^2)^2=\gamma^2$; there are $\binom{n-2}{2}$ such transpositions. Thus
\be
    \sum_{\substack{s\in\calT_1\\ s\neq\tau}} M_{\tau s}(\rho_0) = 2(n-2)\tr(\rho_0^3) + \binom{n-2}{2}\gamma^2 = 2m\,\tr(\rho_0^3) + \frac{m(m-1)}{2}\gamma^2 . \label{eq:appendix_first_shell_moment_estimate}
\ee

Therefore,
\begin{align}
    \E_\text{Haar}\overline{\gamma}_1 &=
    \lim_{m\to -1} \E_{\rm Haar}\overline{\gamma}^{(m)}_1 \nonumber \\
    &= \lim_{m\to -1} d_B^{-(m+1)}\gamma + \frac{1}{d_A}d_B^{-(m+1)} \left(1-\frac{1}{d_B}\right) \left[ 1 + 2m\,\tr(\rho_0^3) +\frac{m(m-1)}{2}\gamma^2 \right] + \calO(d_A^{-2}d_B^{-(m+1)}) \nonumber \\
    &= \gamma + \frac{1}{d_A} \left(1-\frac{1}{d_B}\right) \left[ 1 - 2\,\tr(\rho_0^3) + \gamma^2 \right] + \calO(d_A^{-2}).
\end{align}

Therefore, for the one-step measurement-assisted model, the averaged input sensitivity is

\begin{align}
    \E_{\mathrm{Haar}}\mathcal{S}_1 &= 1 - \E_\text{Haar} \overline{\gamma}_1 \nonumber \\
    &= 1 - \gamma - \frac{1}{d_A} \left(1-\frac{1}{d_B}\right) \left[ 1 - 2\,\tr(\rho_0^3) + \gamma^2 \right] + \calO(d_A^{-2})
\end{align}

\subsection{Multistep Haar twirling and boundary contraction}
\label{app:multistep_concentration}

In the sequential monitored circuit, a Haar-random unitary is applied at each step. Between the input ensemble boundary $\roundbra{\rho_{0}}_A^{\otimes n}\roundbra{0}_B^{\otimes n}$ and the final swap boundary $\roundket{\tau}_A\roundket{D_B^{(n)}}$, we therefore attach $t$ Haar-twirled blocks. Since one layer generates a superposition of permutation sectors, we derive the one-layer rule for a general boundary $\roundket{\alpha}_A\roundket{D_B^{(n)}}$ with $\alpha\in S_n$.

After applying one monitored Haar-twirled block to the general boundary, we obtain
\begin{align}
    \roundbra{0}_B^{\otimes n}\Phi_n\roundket{\alpha}_A\roundket{D_B^{(n)}}
    &= d_B{\rm Wg}(e,d) \sum_{\pi\in S_n} d_A^{\#(\pi^{-1}\alpha)} \roundket{\hat{\pi}}_{A} \nonumber\\
    &\quad + d_B \sum_{\substack{\sigma,\pi\in S_n\\ \sigma\neq\pi}} {\rm Wg}(\sigma^{-1}\pi,d) d_A^{\#(\pi^{-1}\alpha)} \roundket{\hat{\sigma}}_{A}. \label{eq:appendix_monitored_split_alpha}
\end{align}

The diagonal part is organized by the distance from $\alpha$. Denoting $\ell=n-\#(\pi^{-1}\alpha)$, one diagonal term has size
\be
    d_B{\rm Wg}(e,d)d_A^{\#(\pi^{-1}\alpha)} = d_A^{-\ell}d_B^{1-n}\left(1+\calO(d^{-2})\right). \label{eq:appendix_diagonal_power_count_alpha}
\ee
Thus the leading diagonal shell is $\pi=\alpha$, and the first subleading shell is $\pi=\alpha s$ with $s\in\calT_1$:
\be
    d_B{\rm Wg}(e,d) \sum_{\pi\in S_n} d_A^{\#(\pi^{-1}\alpha)} \roundket{\hat{\pi}}_{A}
    =d_B^{1-n}\!\left[\roundket{\alpha}_{A}+\frac{1}{d_A}\sum_{s\in\calT_1}\roundket{\widehat{\alpha s}}_{A}+\calO(d_A^{-2})\right].
    \label{eq:appendix_diagonal_standard_shells_alpha}
\ee

For the off-diagonal part, denote $r=n-\#(\sigma^{-1}\pi)$ and $\ell=n-\#(\pi^{-1}\alpha)$. Then one off-diagonal term has size
\be
    d_B{\rm Wg}(\sigma^{-1}\pi,d)d_A^{\#(\pi^{-1}\alpha)}
    =\calO\!\left(d_A^{-r-\ell}d_B^{1-n-r}\right).
    \label{eq:appendix_offdiagonal_power_count_alpha}
\ee
The largest off-diagonal shell has $r=1$ and $\ell=0$, namely $\pi=\alpha$ and $\sigma=\alpha s$ with $s\in\calT_1$. Keeping this shell gives
\be
    d_B \sum_{\substack{\sigma,\pi\in S_n\\ \sigma\neq\pi}} {\rm Wg}(\sigma^{-1}\pi,d) d_A^{\#(\pi^{-1}\alpha)} \roundket{\hat{\sigma}}_{A}
    =-\frac{1}{d_A}d_B^{-n}\sum_{s\in\calT_1}\roundket{\widehat{\alpha s}}_{A}+\calO(d_A^{-2}d_B^{1-n}).
    \label{eq:appendix_offdiagonal_standard_shell_alpha}
\ee

Combining diagonal and off-diagonal terms through the first subleading order gives the general boundary-connected rule
\begin{align}
    \roundbra{0}_B^{\otimes n} \Phi_n \roundket{\alpha}_A\roundket{D_B^{(n)}}
    &= q_0\roundket{\alpha}_A + q_1 \sum_{s\in\calT_1} \roundket{\widehat{\alpha s}}_A + \calO(d_A^{-2}d_B^{-(m+1)}), \label{eq:appendix_general_first_shell_rule}\\
    \text{where} \quad q_0 &= d_Bd^{-n}d_A^n = d_B^{-(m+1)}, \nonumber\\
    q_1 &= d_B\left[ d^{-n}d_A^{n-1} - d^{-n-1}d_A^n \right] = \frac{1}{d_A}d_B^{-(m+1)} \left(1-\frac{1}{d_B}\right). \nonumber
\end{align}

To derive the multistep evolution, it is therefore useful to introduce the first-shell transfer operator on the full permutation-sector space,
\be
    \mathcal{K} = q_0 I + q_1 A_{\calT_1}, \qquad A_{\calT_1}\roundket{\alpha}_A = \sum_{s\in\calT_1} \roundket{\widehat{\alpha s}}_A . \label{eq:appendix_first_shell_transfer_operator}
\ee
This is the standard first-shell transfer rule on the large permutation-sector space. After $t$ Haar-twirled layers, the backward-propagated boundary from the final swap sector is
\begin{align}
    \mathcal{K}^t\roundket{\tau}_A 
    &= q_0^t \sum_{j=0}^{t} \binom{t}{j} \left(\frac{q_1}{q_0}\right)^j A_{\calT_1}^j\roundket{\tau}_A \nonumber\\
    &= q_0^t \sum_{j=0}^{t} \binom{t}{j} \left(\frac{q_1}{q_0}\right)^j \sum_{s_1,\ldots,s_j\in\calT_1} \roundket{\widehat{\tau s_1\cdots s_j}}_A , \label{eq:appendix_t_step_transfer_expansion}
\end{align}
Attaching the input boundary $\roundbra{\rho_{0}}_A^{\otimes n}$ converts each generated permutation sector into a cycle moment:
\begin{align}
    \E_{\rm Haar}\overline{\gamma}^{(m)}_t
    &= \roundbra{\rho_{0}}_A^{\otimes n} \mathcal{K}^t\roundket{\tau} \nonumber\\
    &= d_B^{-t(m+1)} \sum_{j=0}^{t} \binom{t}{j} \left[ \frac{1}{d_A} \left(1-\frac{1}{d_B}\right) \right]^j \sum_{s_1,\ldots,s_j\in\calT_1} M_{\tau s_1\cdots s_j}(\rho_{0}) + \text{higher Weingarten shells}. \label{eq:appendix_forward_full_series}
\end{align}

For a normalized density matrix $\rho_0$, the permutation overlap satisfies
\begin{equation}
    0\leq M_\sigma(\rho_0)=\prod_{c\in \mathrm{cycles}(\sigma)}\operatorname{Tr}(\rho_0^{|c|})\leq 1 .
\end{equation}
For a fixed $j$, the corresponding contribution is
\begin{equation}
    R_j=d_B^{-t(m+1)}\binom{t}{j}\left[\frac{1}{d_A}\left(1-\frac{1}{d_B}\right)\right]^j\sum_{s_1,\ldots,s_j\in\mathcal T_1}M_{\tau s_1\cdots s_j}(\rho_0).
\end{equation}
Since $|\mathcal T_1|=\binom{n}{2}=\binom{m+2}{2}$, we have
\begin{equation}
    \sum_{s_1,\ldots,s_j\in\mathcal T_1}M_{\tau s_1\cdots s_j}(\rho_0)\leq |\mathcal T_1|^j=\binom{m+2}{2}^j .
\end{equation}
Therefore
\begin{equation}
    R_j\leq d_B^{-t(m+1)}\binom{t}{j}\left[\frac{\binom{m+2}{2}}{d_A}\left(1-\frac{1}{d_B}\right)\right]^j.
\end{equation}
For fixed $d_B$ and $j$, $t \ll d_A$, and using $\binom{t}{j}\leq t^j/j!$, this simplifies to
\begin{equation}
    R_j=\calO\!\left(d_B^{-t(m+1)}\left(\frac{t m^2}{d_A}\right)^j\right).
\end{equation}
and every fixed-$j\geq2$ sector is of order $\calO(d_A^{-j})$. Thus, retaining only the $j=0,1$ sectors is controlled for fixed $d_B$ and $j$ with $t \ll d_A$.

To organize the leading-order terms, let
\be
    r_k \equiv \tr\!\left(\rho_{0}^k\right), \qquad r_1=1, \qquad a \equiv \frac{1}{d_A} \left(1-\frac{1}{d_B}\right). \label{eq:appendix_forward_pk_a_definition}
\ee
At finite integer $m$, with $n=m+2$ and $\tau=(m+1\;m+2)$, define $S_j(m) \equiv \sum_{s_1,\ldots,s_j\in\calT_1} M_{\tau s_1\cdots s_j}(\rho_{0})$. Specifically, the $j=0$ and $j=1$ terms are
\begin{align}
    S_0(m) &\equiv M_\tau(\rho_{0}) = r_2 =\gamma, \label{eq:appendix_forward_S0_finite_m}\\
    S_1(m) &\equiv \sum_{s\in\calT_1} M_{\tau s}(\rho_{0}), \label{eq:appendix_forward_S1_definition}\\
    % S_2(m) &\equiv \sum_{s_1,s_2\in\calT_1} M_{\tau s_1s_2}(\rho_{0}) . \label{eq:appendix_forward_S2_definition}
\end{align}
Then Eq.~\eqref{eq:appendix_forward_full_series} gives, through the first two orders,
\begin{align}
    \E_{\rm Haar}\overline{\gamma}^{(m)}_t
    &= d_B^{-t(m+1)}
    \left[
    S_0(m)+taS_1(m) + \calO(d_A^{-2})%+\binom{t}{2}a^2S_2(m)+\calO(a^3)
    \right] + \text{higher Weingarten shells}. \label{eq:appendix_forward_first_three_shells}
\end{align}

The $j=1$ term is obtained by classifying the relation between $s$ and $\tau$. The term $s=\tau$ gives $\tau s=e$ and contributes $M_e=1$. If $s$ shares one index with $\tau$, there are $2m$ such transpositions and $\tau s$ is a three-cycle, giving $r_3$. If $s$ is disjoint from $\tau$, there are $\binom{m}{2}$ choices and $\tau s$ has two disjoint two-cycles, giving $r_2^2$. Hence
\be
    S_1(m)=1+2mr_3+\binom{m}{2}r_2^2 . \label{eq:appendix_forward_S1_finite_m}
\ee
% For the $j=2$ term, the finite-integer cycle-counting result is
% \begin{align}
% S_2(m) &= \frac{3m^2+13m+2}{2}r_2 + 8m(m-1)p_4 \nonumber\\
% &\quad + 3m(m-1)(m-2)p_3p_2
% + \frac{m(m-1)(m-2)(m-3)}{4}r_2^3 . \label{eq:appendix_forward_S2_finite_m}
% \end{align}

Only after obtaining these finite-$m$ polynomials do we analytically continue to $m\to -1$. The overall prefactor satisfies
\be
    d_B^{-t(m+1)} \longrightarrow 1, \label{eq:appendix_forward_prefactor_replica_limit}
\ee
\begin{align}
    S_0(-1) &= r_2 = \gamma, \nonumber\\
    S_1(-1) &= 1-2r_3+r_2^2 = 1 - 2\tr(\rho_0^3) + \gamma^2 , \nonumber\\
    % S_2(-1) &= -4p_2 + 16p_4 - 18p_2p_3 + 6p_2^3 . \label{eq:appendix_forward_replica_continued_shells}
\end{align}
Therefore,
\begin{align}
    \E_\text{Haar}\overline{\gamma}_t &=
        \lim_{m\to -1} \E_{\rm Haar}\overline{\gamma}^{(m)}_t \nonumber \\
    &= \gamma + \frac{t}{d_A}\left(1-\frac{1}{d_B}\right)\left(1 -2\tr(\rho_0^3) + \gamma^2\right) +\calO(d_A^{-2}) . \label{eq:appendix_forward_replica_limit_to_second_order}
% &\quad + \binom{t}{2}\frac{1}{d_A^2}\left(1-\frac{1}{d_B}\right)^2\left(-4p_2+16p_4-18p_2p_3+6p_2^3\right)
\end{align}

Therefore, for the multistep measurement-assisted model, the averaged input sensitivity is

\begin{align}
    \E_{\mathrm{Haar}}\mathcal{S}_t &= 1 - \E_\text{Haar} \overline{\gamma}_t \nonumber \\
    &= 1 - \gamma - \frac{t}{d_A} \left(1-\frac{1}{d_B}\right) \left[ 1 - 2\,\tr(\rho_0^3) + \gamma^2 \right] + \calO(d_A^{-2})
\end{align}
% For the fully mixed input state, $\rho_{A_0}=I_{A_0}/d_{A_0}$, the moments are $p_k=d_{A_0}^{1-k}$. In particular,
% \be
% 2p_3 = 2\tr\!\left(\rho_{A_0}^3\right) = \frac{2}{d_{A_0}^2}, \label{eq:appendix_forward_fully_mixed_p3}
% \ee
% so this contribution is suppressed as $\calO(d_{A_0}^{-2})$ for large $d_{A_0}$.
\end{widetext}

\section{Circuit Architecture and Monitored Inference Protocol}
\label{app:inference_details}

In this section, we describe the circuit implementations used in the QuDDPM architectures considered in the main text. Both original QuDDPM and truncated QuDDPM are built from two types of circuits: a forward scrambling circuit and a backward denoising circuit. The forward circuit acts only on the system register and generates the sequence of scrambled sample sets. The backward circuit is implemented as a monitored circuit, in which the system is coupled to ancilla qubits, followed by ancilla measurement and reset. In this work, we adopt a circuit structure similar to that in~\cite{zhang2024generative}. Below, we specify the gate structure, ancilla usage, and measurement protocol for these circuits.
\begin{figure}[b]
    \centering
    \includegraphics[width=\columnwidth]{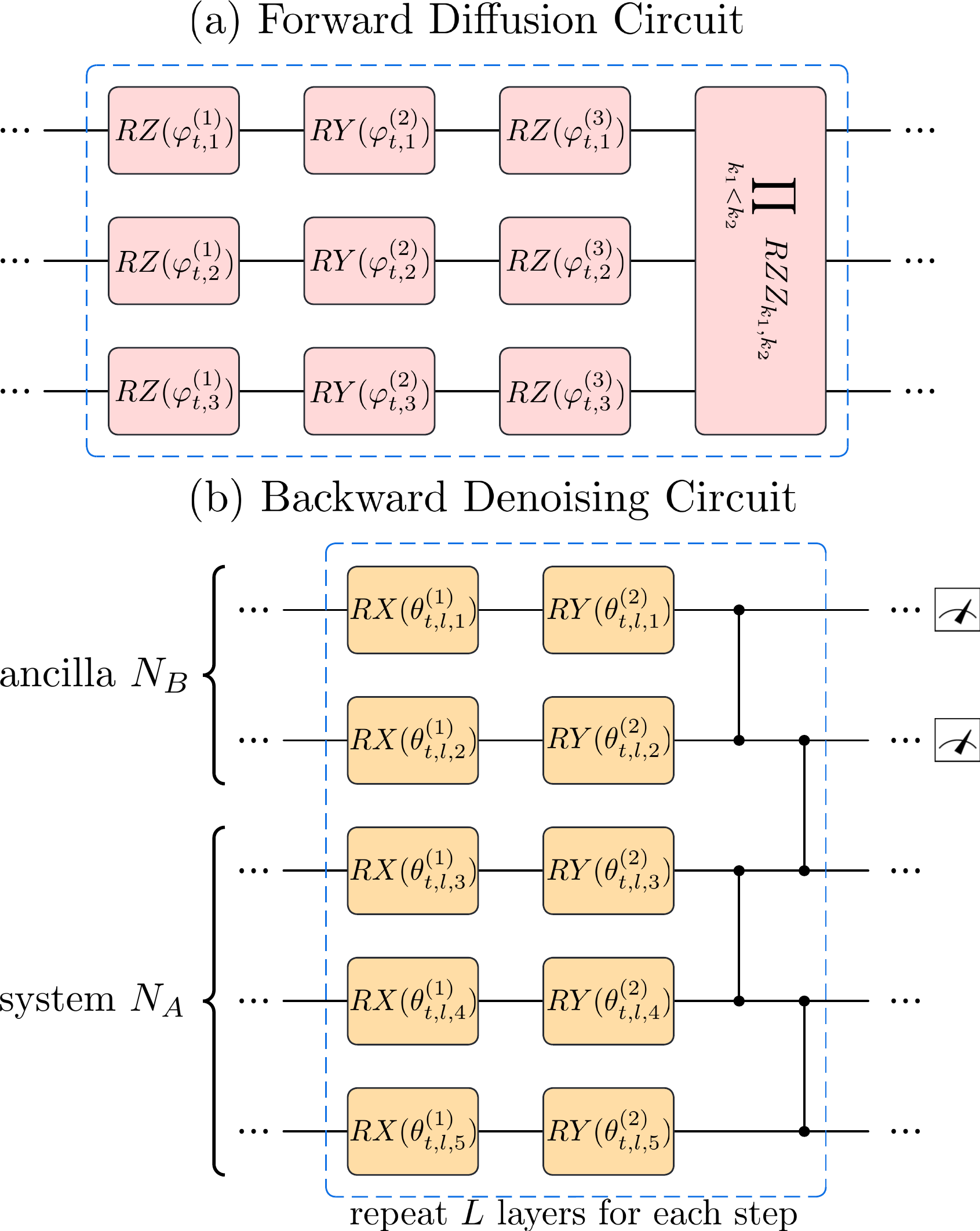}
  \caption{
    Circuit architecture of QuDDPM. (a) Forward diffusion circuit for one diffusion step, with $N_A=3$. (b) One layer of the $L$-layer backward denoising circuit, with $N_A=3$ and $N_B=2$. Here $RX$, $RY$, and $RZ$ are single-qubit Pauli rotations, and $RZZ$ is the two-qubit $Z\otimes Z$ interaction. The entangling gates in (b) are controlled-$Z$ gates.
}
\label{fig:forward_backward_circuit}
\end{figure}

The forward diffusion circuit is implemented using the fast scrambling model~\cite{belyansky2020minimal}. To implement a controllable, gradual diffusion process, the scrambling parameters are scaled according to a diffusion schedule that depends on the system size $N_A$ and the temporal depth $T$. Specifically, as shown in Fig.~\ref{fig:forward_backward_circuit}(a), one block of the forward diffusion circuit $V_t$ is implemented as
\begin{equation}
    V_t(\varphi_t, h_t) = G_{RZZ}(h_t) W_t(\varphi_t)
\end{equation}
where $G_{RZZ}$ consists of a $ZZ$ rotation on every pair of system qubits,
\begin{equation}
    G_{RZZ}(h_t) = \prod_{k_1 <k_2} \text{exp}\left[-i\frac{h_t}{2\sqrt {N_A}} Z_{k_1}Z_{k_2} \right],
\end{equation}
and $W$ applies single-qubit rotations to every system qubit,
\begin{align}
    W(\varphi_t) = \bigotimes_{k=1}^{N_A}\text{exp}\left[-\frac{i}{2}\left(\varphi_{t,k}^{(3)} Z_k+ \varphi_{t,k}^{(2)} Y_k+ \varphi_{t,k}^{(1)} Z_k\right)\right].
\end{align}
To generate the scrambled target ensemble $\calE_t$, the model applies the forward scrambling blocks from step $1$ through step $t$ to the target ensemble $\calE_\text{tar}$. The scrambling parameters $h_t$ and $\varphi$ are randomly sampled and transformed according to the tunable diffusion schedule, allowing control over the step-by-step diffusion speed from the input ensemble $\calE_\text{tar}$ toward the Haar-random ensemble $\calE_\text{Haar}$.

The backward denoising circuit is implemented with a measurement-assisted generative circuit, as described in Sec.~\ref{subsec:inference_circuit} and Ref.~\cite{zhang2024generative}. In this work, the global unitary $\tilde{U}_{t}(\theta_t)$ across the system and ancilla is implemented using a parameterized, hardware-efficient ansatz~\cite{kandala2017hardware}. To simplify circuit deployment, we assume nearest-neighbor connectivity for the entangling gates. Specifically, 
\begin{equation}
    \tilde{U}_t(\theta_t) = \prod_{l=1}^{L}\tilde{G}_{CZ} \tilde{W}_t(\theta_{t,l}),
\end{equation}
where
\begin{align}
    &\tilde{G}_{CZ} = \bigotimes_{k=1}^{\lfloor(N-1)/2\rfloor} CZ_{2k, 2k+1} \; \bigotimes_{k=1}^{\lfloor N/2\rfloor} CZ_{2k-1, 2k} \\
    &\tilde W_t(\theta_{t, l})= \bigotimes_{k=1}^{N}\text{exp}\left[-\frac{i}{2}\left(\theta_{t,l, k}^{(2)} Y_k+ \theta_{t,l,k}^{(1)} X_k\right)\right]
\end{align}
and $N=N_A + N_B$ is the total number of qubits, while $CZ$ denotes the controlled-$Z$ gate. The choice of entangling gate can be adapted to the native gate set and connectivity constraints of the quantum device. At each backward step, the circuit repeats $L$ layers of the backward circuit block above and then performs a computational-basis measurement on the ancilla. Before proceeding to the next denoising step, the model resets the ancilla qubits to $\ket{0}_B^{\otimes N_B}$.

\section{Numerical simulation of random conditional transport}

In Sec.~\ref{sec:concentration}, we present numerical simulation results of conditional transport with Haar-random unitary at each step. We simulate a system register $A$ with $N_A$ qubits and an ancilla register $B$ with $N_B$ qubits. For typical random dynamics, each step is implemented by an independent Haar-random unitary $U_t$ followed by mid-circuit measurement on the ancilla qubits.
% ,
% \begin{equation}
%     U_t \in \mathrm{U}(d_A d_B), \qquad t=0,\ldots,T-1 .
% \end{equation}
% At each step, the current state on $AB$ is embedded as $|0\rangle\!\langle 0|_B\otimes \rho_{A,t}$, evolved by $U_t$, and then subjected to a projective measurement of $B$ in the computational basis. 
For measurement outcome $\bmz_t\in\{0,\ldots,d_B-1\}$, the conditional update follows Eq.~\eqref{eq:cond_ensemble_def}.
Iterating this rule gives a conditional trajectory $\boldsymbol{\bmz}=(\bmz_0,\ldots,\bmz_{T-1})$.

To estimate input sensitivity, we initialize the input sample set $S_0$ in its density-operator form $\rho_{S_0}$ by averaging over $10^4$ Haar-random pure states on $A$, which is nearly a maximally mixed state $\rho_{S_0}\simeq I_A/d_A$. We then sample measurement trajectories by Monte Carlo, drawing each $\bmz_t$ from its Born probability $p(\bmz_t|\rho_{A,t-1})$. For each trajectory, we record the purity of the average state of the final output state ensemble at step $t$:
\begin{equation}
    \gamma_t(\boldsymbol{\bmz})=\tr\!\left(\rho_{A,t|\boldsymbol{\bmz}}^2\right).
\end{equation}
For the main numerical data in Figs.~\ref{fig:concentration_numerics}(a) and~\ref{fig:concentration_numerics}(e), we randomly sample $10^3$ trajectories for $N_A=1,\ldots,6$ and $N_B=1,2$ to obtain an estimation on the average behavior. For the trained-model inset, we use $10^4$ trajectories.

The numerical concentration time in Fig.~\ref{fig:concentration_numerics}(e) is extracted from the input-sensitivity curve using a threshold criterion. For a fixed tolerance $\epsilon$, we define the sample-averaged concentration time as
\begin{equation}
    \widehat{\tau}_\epsilon=\min\{t:\widehat{\mathcal{S}}_t\le \epsilon\}=\min\{t:\widehat{\gamma}_t\ge 1-\epsilon\}.
\end{equation}
We compare these estimates with the theoretical concentration-time curve from Eq.~\eqref{eq:concentration_time}.

For the conditional-transport Bloch-sphere plots in Figs.~\ref{fig:concentration_numerics}(b) and~\ref{fig:concentration_numerics}(d), we sample an ensemble of $N_{\rm data}=100$ Haar-random pure input states and choose a small number of fixed trajectories, typically $N_{\rm traj}=15$. For each trajectory $\boldsymbol{\bmz}$, every input state is propagated through the same Haar-random unitary sequence with the prescribed measurement outcomes. At time $t$, each conditional pure state $\ket{\psi_{k|\boldsymbol{\bmz}}}_{A_t}$ is plotted using its Bloch coordinates
\begin{equation}
    \left(\langle X\rangle,\langle Y\rangle,\langle Z\rangle\right)=\left(\langle\psi|X|\psi\rangle,\langle\psi|Y|\psi\rangle,\langle\psi|Z|\psi\rangle\right).
\end{equation}
For the logical space spanned by the repetition code, we choose
\begin{align}
     &Z_L=\begin{cases} Z^{\otimes N_A}, &\text{for odd } N_A,\\ Z\otimes I^{\otimes(N_A-1)}, & \text{otherwise}, \end{cases} \\ 
     &X_L=X^{\otimes N_A}, \quad Y_L=iX_LZ_L.
\end{align}
The plotted coordinates are therefore
\begin{equation}
    \left(\langle X_L\rangle,\langle Y_L\rangle,\langle Z_L\rangle\right).
\end{equation}

\section{Numerical Training Protocol}

The general training paradigm is adopted directly from~\cite{zhang2024generative} and is depicted in Fig.~\ref{fig:truncated_QuDDPM}. To evaluate the trade-off among accuracy, generative power, and input sensitivity, we vary the temporal depth in the original and truncated QuDDPM. For the original QuDDPM, we adjust the diffusion schedule according to the total depth to ensure a smooth interpolation from $\calE_\text{gen}$ to $\calE_\text{Haar}$, while for truncated QuDDPM, the diffusion schedule is set in advance and fixed across different $T_\text{trun}$ experiments. 
For all numerical simulations, original and truncated QuDDPM models are trained using the same optimization settings unless explicitly stated otherwise. We use the Adam optimizer with learning rate $\eta=0.001$. The target and generated sample sets each contain $N_{\rm tar}=N_{\rm gen}=5000$ samples, and at each training step, we set the batch size to $B=500$ to estimate the Wasserstein distance. Circuit parameters are initialized randomly from a Gaussian distribution. Early stopping is applied after $60\%$ of the prescribed 3000 epochs: if the training loss does not improve within the subsequent 200-epoch patience window, the optimization is terminated.

Our numerical implementation uses TensorCircuit-NG~\cite{zhang2023tensorcircuit}
to simulate the monitored circuit dynamics and JAX~\cite{jax2018github} automatic differentiation to optimize the variational parameters. The Wasserstein distance in the training objective is approximated by the Sinkhorn distance~\cite{cuturi2013Sinkhorn} computed with the OTT~\cite{cuturi2022optimal} package.
% \BZ{cite the library for circuit simulation.}

% Bibliography
\bibliography{myref}

@misc{Github,
  author = {Mo, Runzhe and Zhang, Bingzhi and Zhuang, Quntao},
  title = {Data and code for paper 'Measurement-induced overconcentration in quantum generative models'},
  year = {2026},
  howpublished = {\url{https://github.com/hairlessdevil/measurement_induced_overconcentration.git}}
}

@article{du2022efficient,
title = {Efficient Measure for the Expressivity of Variational Quantum Algorithms},
author = {Du, Yuxuan and Tu, Zhuozhuo and Yuan, Xiao and Tao, Dacheng},
journal = {Physical Review Letters},
volume = {128},
number = {8},
pages = {080506},
year = {2022},
doi = {10.1103/PhysRevLett.128.080506}
}

@article{sim2019expressibility,
title = {Expressibility and Entangling Capability of Parameterized Quantum Circuits for Hybrid Quantum-Classical Algorithms},
author = {Sim, Sukin and Johnson, Peter D. and Aspuru-Guzik, Al{'a}n},
journal = {Advanced Quantum Technologies},
volume = {2},
number = {12},
pages = {1900070},
year = {2019},
doi = {10.1002/qute.201900070}
}

@article{mcclean2018barren,
title = {Barren Plateaus in Quantum Neural Network Training Landscapes},
author = {McClean, Jarrod R. and Boixo, Sergio and Smelyanskiy, Vadim N. and Babbush, Ryan and Neven, Hartmut},
journal = {Nature Communications},
volume = {9},
number = {1},
pages = {4812},
year = {2018},
doi = {10.1038/s41467-018-07090-4}
}

@article{cerezo2021cost,
title = {Cost Function Dependent Barren Plateaus in Shallow Parametrized Quantum Circuits},
author = {Cerezo, M. and Sone, Akira and Volkoff, Tyler and Cincio, Lukasz and Coles, Patrick J.},
journal = {Nature Communications},
volume = {12},
number = {1},
pages = {1791},
year = {2021},
doi = {10.1038/s41467-021-21728-w}
}

@article{zhang2025energy,
  title={Energy-dependent barren plateau in bosonic variational quantum circuits},
  author={Zhang, Bingzhi and Zhuang, Quntao},
  journal={Quantum Science and Technology},
  volume={10},
  number={1},
  pages={015009},
  year={2025},
  publisher={IOP Publishing}
}

@article{lu2020quantum,
title = {Quantum Adversarial Machine Learning},
author = {Lu, Sirui and Duan, Lu-Ming and Deng, Dong-Ling},
journal = {Physical Review Research},
volume = {2},
number = {3},
pages = {033212},
year = {2020},
doi = {10.1103/PhysRevResearch.2.033212}
}

@article{liao2021robust,
title = {Robust in Practice: Adversarial Attacks on Quantum Machine Learning},
author = {Liao, Haoran and Convy, Ian and Huggins, William J. and Whaley, K. Birgitta},
journal = {Physical Review A},
volume = {103},
number = {4},
pages = {042427},
year = {2021},
doi = {10.1103/PhysRevA.103.042427}
}

@article{kwun2025mixed,
  title={Mixed-state quantum denoising diffusion probabilistic model},
  author={Kwun, Gino and Zhang, Bingzhi and Zhuang, Quntao},
  journal={Phys. Rev. A},
  volume={111},
  number={3},
  pages={032610},
  year={2025},
  doi={10.1103/PhysRevA.111.032610},
  url={https://link.aps.org/doi/10.1103/PhysRevA.111.032610},
  publisher={APS}
}

@article{marrero2021entanglement,
  title={Entanglement-induced barren plateaus},
  author={Ortiz Marrero, Carlos and Kieferov{\'a}, M{\'a}ria and Wiebe, Nathan},
  journal={PRX Quantum},
  volume={2},
  number={4},
  pages={040316},
  year={2021},
  doi={10.1103/PRXQuantum.2.040316},
  url={https://link.aps.org/doi/10.1103/PRXQuantum.2.040316},
  publisher={APS}
}

@article{parigi2025quantum,
  title={Quantum-Noise-Driven Generative Diffusion Models},
  author={Parigi, Marco and Martina, Stefano and Caruso, Filippo},
  journal={Adv. Quantum Technol.},
  volume={8},
  number={12},
  pages={2300401},
  year={2025},
  url={https://arxiv.org/abs/2308.12013},
  publisher={Wiley Online Library}
}

@article{yao2026hierarchy,
  title={Hierarchy of discriminative power and complexity in learning quantum ensembles},
  author={Yao, Jian and Li, Pengtao and Chen, Xiaohui and Zhuang, Quntao},
  journal={arXiv:2601.22005},
  url={https://arxiv.org/abs/2601.22005},
  year={2026}
}

@article{ho2022exact,
  title={Exact emergent quantum state designs from quantum chaotic dynamics},
  author={Ho, Wen Wei and Choi, Soonwon},
  journal={Phys. Rev. Lett.},
  volume={128},
  number={6},
  pages={060601},
  year={2022},
  doi={10.1103/PhysRevLett.128.060601},
  url={https://link.aps.org/doi/10.1103/PhysRevLett.128.060601},
  publisher={APS}
}

@article{kandala2017hardware,
  title={Hardware-efficient variational quantum eigensolver for small molecules and quantum magnets},
  author={Kandala, Abhinav and Mezzacapo, Antonio and Temme, Kristan and Takita, Maika and Brink, Markus and Chow, Jerry M and Gambetta, Jay M},
  journal={Nature},
  volume={549},
  number={7671},
  pages={242--246},
  year={2017},
  doi={10.1038/nature23879},
  url={https://doi.org/10.1038/nature23879},
  publisher={Nature Publishing Group UK London}
}

@inproceedings{Kingma2014Auto-Encoding,
    author    = {Kingma, Diederik P. and Welling, Max},
    title     = {Auto-Encoding Variational Bayes},
    booktitle = {Proceedings of the 2nd International Conference on Learning Representations},
    address   = {Banff, Canada},
    year      = {2014},
    url       = {https://openreview.net/forum?id=33X9fd2-9FyZd}
}

@inproceedings{Rezende2015Variational,
    author    = {Rezende, Danilo and Mohamed, Shakir},
    title     = {Variational Inference with Normalizing Flows},
    booktitle = {Proceedings of the 32nd International Conference on Machine Learning},
    publisher = {PMLR},
    address   = {Lille, France},
    year      = {2015},
    volume    = {37},
    pages     = {1530--1538},
    url       = {https://proceedings.mlr.press/v37/rezende15.html}
}

@inproceedings{song2021scorebased,
	author = {Yang Song and Jascha Sohl-Dickstein and Diederik P Kingma and Abhishek Kumar and Stefano Ermon and Ben Poole},
	booktitle = {The Ninth International Conference on Learning Representations},
	title = {Score-Based Generative Modeling through Stochastic Differential Equations},
	url = {https://openreview.net/forum?id=PxTIG12RRHS},
	year = {2021}}

@article{belyansky2020minimal,
    title={Minimal Model for Fast Scrambling},
    author={Belyansky, Ron and Bienias, Przemyslaw and Kharkov, Yaroslav A and Gorshkov, Alexey V and Swingle, Brian},
    journal={Phys. Rev. Lett.},
    volume={125},
    number={13},
    pages={130601},
    year={2020},
    doi={10.1103/PhysRevLett.125.130601},
    url={https://link.aps.org/doi/10.1103/PhysRevLett.125.130601},
    publisher={APS}
}

@article{dallaire2018quantum,
  title={Quantum generative adversarial networks},
  author={Dallaire-Demers, Pierre-Luc and Killoran, Nathan},
  journal={Phys. Rev. A},
  volume={98},
  number={1},
  pages={012324},
  year={2018},
  doi={10.1103/PhysRevA.98.012324},
  url={https://link.aps.org/doi/10.1103/PhysRevA.98.012324},
  publisher={APS}
}

@article{lloyd2018quantum,
  title={Quantum generative adversarial learning},
  author={Lloyd, Seth and Weedbrook, Christian},
  journal={Phys. Rev. Lett.},
  volume={121},
  number={4},
  pages={040502},
  year={2018},
  doi={10.1103/PhysRevLett.121.040502},
  url={https://link.aps.org/doi/10.1103/PhysRevLett.121.040502},
  publisher={APS}
}

@article{zhang2023tensorcircuit,
  title={Tensorcircuit: a quantum software framework for the nisq era},
  author={Zhang, Shi-Xin and Allcock, Jonathan and Wan, Zhou-Quan and Liu, Shuo and Sun, Jiace and Yu, Hao and Yang, Xing-Han and Qiu, Jiezhong and Ye, Zhaofeng and Chen, Yu-Qin and others},
  journal={Quantum},
  volume={7},
  pages={912},
  year={2023},
  doi={10.22331/q-2023-02-02-912},
  url={https://doi.org/10.22331/q-2023-02-02-912},
  publisher={Verein zur F{\"o}rderung des Open Access Publizierens in den Quantenwissenschaften}
}

@article{coyle2020born,
  title={The Born supremacy: quantum advantage and training of an Ising Born machine},
  author={Coyle, Brian and Mills, Daniel and Danos, Vincent and Kashefi, Elham},
  journal={npj Quantum Inf.},
  volume={6},
  number={1},
  pages={60},
  year={2020},
  doi={10.1038/s41534-020-00288-9},
  url={https://doi.org/10.1038/s41534-020-00288-9},
  publisher={Nature Publishing Group UK London}
}

@article{liu2018differentiable,
  title={Differentiable learning of quantum circuit born machines},
  author={Liu, Jin-Guo and Wang, Lei},
  journal={Phys. Rev. A},
  volume={98},
  number={6},
  pages={062324},
  year={2018},
  doi={10.1103/PhysRevA.98.062324},
  url={https://link.aps.org/doi/10.1103/PhysRevA.98.062324},
  publisher={APS}
}

@article{zhang2025scaling,
  title = {Scaling Laws of Quantum Information Lifetime in Monitored Quantum Dynamics},
  author = {Zhang, Bingzhi and Hu, Fangjun and Mo, Runzhe and Chen, Tianyang and T\"ureci, Hakan E. and Zhuang, Quntao},
  journal = {Phys. Rev. X},
  volume = {16},
  issue = {2},
  pages = {021027},
  numpages = {53},
  year = {2026},
  month = {May},
  publisher = {American Physical Society},
  doi = {10.1103/7717-1mw2},
  url = {https://link.aps.org/doi/10.1103/7717-1mw2}
}

@article{zhang2024generative,
  title={Generative quantum machine learning via denoising diffusion probabilistic models},
  author={Zhang, Bingzhi and Xu, Peng and Chen, Xiaohui and Zhuang, Quntao},
  journal={Phys. Rev. Lett.},
  volume={132},
  number={10},
  pages={100602},
  year={2024},
  doi={10.1103/PhysRevLett.132.100602},
  url={https://link.aps.org/doi/10.1103/PhysRevLett.132.100602},
  publisher={APS}
}

@article{zhang2025holographic,
  title   = {Holographic deep thermalization for secure and efficient quantum random state generation},
  author  = {Zhang, Bingzhi and Xu, Peng and Chen, Xiaohui and Zhuang, Quntao},
  journal = {Nat. Commun.},
  volume  = {16},
  pages   = {6341},
  year    = {2025},
  url     = {https://arxiv.org/abs/2411.03587},
}

@article{cao2025mitigating,
  title         = {Mitigating Barren plateaus in quantum denoising diffusion probabilistic models},
  author        = {Haipeng Cao and Kaining Zhang and Dacheng Tao and Zhaofeng Su},
  year          = {2025},
  journal       = {arXiv:2512.06695},
  url           = {https://arxiv.org/abs/2512.06695}
}

@article{cacioppo2023quantum,
  title = {Quantum Diffusion Models},
  author = {Cacioppo, Andrea and Colantonio, Lorenzo and Bordoni, Simone and Giagu, Stefano},
  journal = {arXiv:2311.15444},
  year = {2023},
  url = {https://arxiv.org/abs/2311.15444}
}

@article{ma2025quantum,
  author  = {Ma, QuanGong and Hao, ChaoLong and Si, NianWen and Chen, Geng and Zhang, Jiale and Qu, Dan},
  title   = {Quantum adversarial generation of high-resolution images},
  journal = {EPJ Quantum Technol.},
  volume  = {12},
  pages   = {3},
  year    = {2025},
  doi     = {10.1140/epjqt/s40507-024-00304-3}
}

@inproceedings{goodfellow2014generative,
  title = {Generative Adversarial Nets},
  author = {Goodfellow, Ian and Pouget-Abadie, Jean and Mirza, Mehdi and Xu, Bing and Warde-Farley, David and Ozair, Sherjil and Courville, Aaron and Bengio, Yoshua},
  booktitle = {Advances in Neural Information Processing Systems},
  year = {2014},
  publisher = {Curran Associates, Inc.},
  volume = {27},
  url = {https://proceedings.neurips.cc/paper_files/paper/2014/file/f033ed80deb0234979a61f95710dbe25-Paper.pdf},
}

@inproceedings{oord2016pixel,
    title = {Pixel Recurrent Neural Networks},
    author = {van den Oord, A{\"a}ron and Kalchbrenner, Nal and Kavukcuoglu, Koray},
    booktitle = {Proceedings of the 33rd International Conference on Machine Learning},
    pages = {1747--1756},
    year = {2016},
    volume = {48}, 
    address = {New York, NY},
    publisher = {PMLR},
    url = {https://proceedings.mlr.press/v48/oord16.html}
}

@inproceedings{ho2020denoising,
    title={Denoising diffusion probabilistic models},
    author={Ho, Jonathan and Jain, Ajay and Abbeel, Pieter},
    booktitle={Advances in Neural Information Processing Systems},
    volume={33},
    pages={6840--6851},
    year={2020},
    publisher = {Curran Associates, Inc.},
    url = {https://proceedings.neurips.cc/paper/2020/hash/4c5bcfec8584af0d967f1ab10179ca4b-Abstract.html}
}

@inproceedings{Lipman2023Flow,
  author    = {Lipman, Yaron and Chen, Ricky T. Q. and Ben-Hamu, Heli and Nickel, Maximilian and Le, Matt},
  title     = {Flow Matching for Generative Modeling},
  booktitle = {Proceedings of the Eleventh International Conference on Learning Representations},
  address   = {Kigali, Rwanda},
  year      = {2023},
  url       = {https://openreview.net/forum?id=PqvMRDCJT9t}
}

@inproceedings{cuturi2013Sinkhorn,
  author    = {Marco Cuturi},
  title     = {Sinkhorn Distances: Lightspeed Computation of Optimal Transport},
  booktitle = {Advances in Neural Information Processing Systems 26},
  pages     = {2292--2300},
  publisher = {Curran Associates, Inc.},
  year      = {2013},
  url       = {https://proceedings.neurips.cc/paper_files/paper/2013/file/af21d0c97db2e27e13572cbf59eb343d-Paper.pdf}
}

@inproceedings{heusel2017gans,
  title     = {GANs Trained by a Two Time-Scale Update Rule Converge to a Local Nash Equilibrium},
  author    = {Heusel, Martin and Ramsauer, Hubert and Unterthiner, Thomas and Nessler, Bernhard and Hochreiter, Sepp},
  booktitle = {Advances in Neural Information Processing Systems},
  volume    = {30},
  year      = {2017},
  url       = {https://arxiv.org/abs/1706.08500}
}

@inproceedings{binkowski2018demystifying,
  title     = {Demystifying MMD GANs},
  author    = {Bi{\'n}kowski, Miko{\l}aj and Sutherland, Danica J. and Arbel, Michael and Gretton, Arthur},
  booktitle = {International Conference on Learning Representations},
  year      = {2018},
  url       = {https://arxiv.org/abs/1801.01401}
}

@inproceedings{zhu2017toward,
  title     = {Toward Multimodal Image-to-Image Translation},
  author    = {Zhu, Jun-Yan and Zhang, Richard and Pathak, Deepak and Darrell, Trevor and Efros, Alexei A. and Wang, Oliver and Shechtman, Eli},
  booktitle = {Advances in Neural Information Processing Systems},
  volume    = {30},
  year      = {2017},
  url       = {https://arxiv.org/abs/1711.11586}
}

@inproceedings{yang2019diversity,
  title     = {Diversity-Sensitive Conditional Generative Adversarial Networks},
  author    = {Yang, Dingdong and Hong, Seunghoon and Jang, Yunseok and Zhao, Tianchen and Lee, Honglak},
  booktitle = {International Conference on Learning Representations},
  year      = {2019},
  url       = {https://arxiv.org/abs/1901.09024}
}

@inproceedings{mao2019mode,
  title     = {Mode Seeking Generative Adversarial Networks for Diverse Image Synthesis},
  author    = {Mao, Qi and Lee, Hsin-Ying and Tseng, Hung-Yu and Ma, Siwei and Yang, Ming-Hsuan},
  booktitle = {Proceedings of the IEEE/CVF Conference on Computer Vision and Pattern Recognition},
  pages     = {1429--1437},
  year      = {2019},
  url       = {https://arxiv.org/abs/1903.05628}
}

@inproceedings{chen2016infogan,
  title     = {InfoGAN: Interpretable Representation Learning by Information Maximizing Generative Adversarial Nets},
  author    = {Chen, Xi and Duan, Yan and Houthooft, Rein and Schulman, John and Sutskever, Ilya and Abbeel, Pieter},
  booktitle = {Advances in Neural Information Processing Systems},
  volume    = {29},
  year      = {2016},
  url       = {https://arxiv.org/abs/1606.03657}
}

@misc{jax2018github,
  author = {James Bradbury and Roy Frostig and Peter Hawkins and Matthew James Johnson and Yash Katariya and Chris Leary and Dougal Maclaurin and George Necula and Adam Paszke and Jake Vander{P}las and Skye Wanderman-{M}ilne and Qiao Zhang},
  title = {{JAX}: composable transformations of {P}ython+{N}um{P}y programs},
  url = {http://github.com/jax-ml/jax},
  year = {2018}
}

@article{cuturi2022optimal,
  title = {Optimal Transport Tools (OTT): A JAX Toolbox for all things Wasserstein},
  author = {Cuturi, Marco and Meng-Papaxanthos, Laetitia and Tian, Yingtao and Bunne, Charlotte and Davis, Geoff and Teboul, Olivier},
  journal = {arXiv:2201.12324},
  year = {2022},
  url = {https://arxiv.org/abs/2201.12324}
}

\end{document}